\documentclass[aps,prb,showpacs,reprint,unsortedaddress]{revtex4-1}

\usepackage{hyperref}
\usepackage{array}
\usepackage{dcolumn}
\usepackage{graphicx}
\usepackage{color}
\usepackage{amsmath}
\usepackage{amssymb}
\usepackage{amsfonts}
\usepackage{dsfont}
\usepackage{epstopdf}
\DeclareMathOperator{\Tr}{Tr}

\newcommand\ket[1]{|#1\rangle}

\begin{document}

\title{Finite temperature dynamical correlations for the dimerized spin-1/2 chain}

\author{E. Coira}
\affiliation{DQMP, University of Geneva, 24 Quai Ernest Ansermet, 1211 Geneva,  Switzerland}
\author{P. Barmettler}
\affiliation{SPOUD AG, Effingerstrasse 23, 3008 Bern, Switzerland}
\author{T. Giamarchi}
\affiliation{DQMP, University of Geneva, 24 Quai Ernest Ansermet, 1211 Geneva,  Switzerland}
\author{C. Kollath}
\affiliation{HISKP, University of Bonn, Nussallee 14-16, 53115 Bonn, Germany}

\begin{abstract}
We use the density matrix renormalization group method (DMRG) to compute the frequency and momentum resolved spin-spin correlation functions of a dimerized spin-1/2 chain under a magnetic field at finite temperature. The spectral features strongly depend on the regime of the magnetic field. For increasing magnetic fields, the transitions from a gapped spin liquid phase to a Tomonaga-Luttinger liquid, and then to a totally polarized phase, can be identified in the spectra. Compared to the zero temperature case, the finite temperature excitations give rise to additional spectral features that we compute numerically and identify analytically as transitions from thermally excited states. We compute quantitatively the broadening of the dispersion of a single spin-flip excitation due to the temperature and find a strong asymmetric broadening. We discuss the consequences of these findings for neutron experiments on dimerized one dimensional quantum chains.
\end{abstract}

\maketitle

\section{Introduction} \label{sec:introduction}

Quantum magnets have proven to be extremely rich physical systems with a large variety of physical behaviors, ranging from quantum solids, for which the system is magnetically ordered, to quantum liquids for which the spin-spin correlation functions decay rapidly\cite{Auerbach_book_spins}. In addition to the large number of models that can be studied analytically or numerically, depending on the dimensions and on the nature of the spin or the lattice, some experimental realizations are nowadays available, allowing for a fruitful exchange between theory and experiments.

One of the important characteristics of quantum magnets is that usually the microscopic Hamiltonian describing such systems is quite well known\cite{giamarchi_book_1d,mikeskaKolezhuk2004}. This is at variance with their itinerant counterpart, for which the difficult problem of the screening of the Coulomb interaction leads to Hamiltonians which are difficult to be determined quantitatively. In these cases caricatural descriptions with models such as the Hubbard model must be made. For quantum magnets the excellent knowledge of the Hamiltonians as well as the various precise probes such as neutron scattering experiments\cite{lovesey_neutron_scattering}, NMR\cite{abragam_rmn,slichter_rmn}, etc., giving direct access to the spin-spin correlations of the system, open the possibility to make very precise comparisons between the theoretical description, if available, and experiments. It also allows in principle to use such systems as quantum simulators for itinerant bosonic systems\cite{ward_simulator_TLL_review}.

However, in order to do such comparisons it is important to have powerful methods that allow to obtain the physical observables directly from the Hamiltonian without having to use uncontrolled approximations. This is especially true if one looks at systems with moderately large magnetic exchange constants such that they can be manipulated by experimentally achievable magnetic fields. In that case temperature and magnetic exchange are not as separated as in the usual norm for magnetic systems, making methods such as a brute force application of field theory less accurate. Implementing this program is difficult, but in particular in one dimension it is possible to use a combination of field theory methods as well as numerical ones to make quantitative predictions.

One of the methods that has proven to be particularly useful is the time-dependent Density Matrix Renormalization Group (DMRG)\cite{Vidal2004,WhiteFeiguin2004,daleyVidal2004}, which has allowed to obtain the frequency and momentum resolved spin-spin correlation functions for several one dimensional systems\cite{bouillot_dynamical_ladder}. Compared to other methods such as Quantum Monte-Carlo (QMC) it has the advantage to give direct access to the real time dynamics and to avoid the very delicate problem of the analytic continuation from imaginary time. However, due to the numerical costliness, these predictions were limited to zero temperature, and only finite temperature calculations of the thermodynamics were accessible\cite{Zwolak_finiteT_DMRG, Verstraete_finiteT_DMRG,feiguin_finite_temp_spinchains}. Recently, full dynamical calculations at finite temperature\cite{karrasch_tdmrg_finiteT,barthel_tdmrg_finiteT}, in the framework of time dependent matrix product states reformulation of DMRG\cite{Schollwoeck2011}, have been performed in simple cases, opening the path to the study of thermal effects on spectral functions. Such studies were for example performed for the nuclear magnetic resonance relaxation in quantum spin-1/2 chains \cite{coira_nmr,dupont_nmr} or for the quantum critical point corresponding to magnetization saturation
\cite{barthel_qcpb,blosser_qcp}.

In this paper we use such methods to make an analysis of the properties of a dimerized system under magnetic field. Dimerized systems have, in the absence of magnetic field, a ground state that is a spin liquid with a finite spin gap and interesting features as multiparticle continua. Bound states have been found in their spectra at zero temperature \cite{barnes_dimerized,uhrig_dimerized,brooksharris_dim,zhengsingh2006,bouzerar_breathers,trebstsingh2000,schmidtuhrig2003,schmidtuhrig2004,collinszheng2006, hamersingh2003,essler97_ff_sg}. Application of the magnetic field closes the gap and transforms the system into a Tomonaga-Luttinger liquid (TLL)\cite{giamarchi_book_1d} in a similar way than for ladder systems\cite{chitra_spinchains_field}. At low energies, and if the dimerization is small compared to the main exchange, such systems can be very well described by a sine-Gordon theory \cite{giamarchi_book_1d} and the closure of the gap by the magnetic field is in the universality class of commensurate-incommensurate (or Pokrovsky-Talapov \cite{pokrovsky_talapov_prl}) universality class. Although the sine-Gordon theory or the mapping to the TLL allows to take partly the effects of the temperature on the correlation functions into account, this is clearly limited to energies or temperatures much smaller than the average exchange. In order to go beyond this limitation the direct numerical approach becomes necessary.

One quantity which is particularly difficult to estimate analytically is the thermal broadening of the modes appearing in the correlation functions. For ladders, estimations based on bond-operators description have been performed\cite{normand_bond_spinladder} in the gapped regime. Similar analyses are in principle directly applicable to the dimerized systems as well. However, these approximations mostly find a shift of an otherwise perfectly sharp spectrum. Semi-classical treatments \cite{damle_ladder} were used to compute the thermal broadening for antiferromagnetic gapped Heisenberg chains. More recently for dimerized systems, exact diagonalization for small systems \cite{mikeskaluckmann2006} and form factors expansions \cite{james_thermbroad_dimer} showed an asymmetric non-gaussian broadening at zero magnetic field. Such asymmetric broadening was also observed experimentally in gapped 3D antiferromagnets\cite{quinterocastro_broadening} and in dimerized chains in the material $\left[\text{Cu}(\text{NO}_3)_2\cdot2.5\text{D}_2\text{O}\right]$\cite{groitlhabicht2016} but full determinations of the thermal effects remains elusive.

We will thus here use a finite temperature DMRG method to investigate the various spin-spin correlations in such dimerized systems and in particular compute the detailed thermal broadening of their most prominent modes. We indeed find an asymmetric broadening of these modes and determine its temperature dependence in an essentially exact way from the numerical data.

The plan of the paper is as follows. In Section~\ref{sec:models} we summarize the main features of the model by discussing the Hamiltonian and its corresponding phase diagram. Section~\ref{sec:dmrg} briefly presents the algorithm adopted for the numerical computation. In Section~\ref{sec:procedure} we describe in detail the correlations that we compute, how, and its relation with neutron scattering measurements. We then move in Section~\ref{sec:spectra} to the results for dynamical correlations for the different values of temperature and field considered, and we give an interpretation of the different structures seen. In addition, we show how temperature affects the characteristics of the spectra. Section~\ref{sec:conclusions} presents conclusions and perspectives. In the Appendix we start by showing some convergence checks for the numerics (\ref{app:convergence}), then in \ref{app:strongbondpicture} and \ref{app:lowenstru} we give details about some calculations made to interpret some of the structures seen in the intensity plots for correlations.

\section{The model} \label{sec:models}

\begin{figure}
  \centering
  \includegraphics[width=0.48\textwidth]{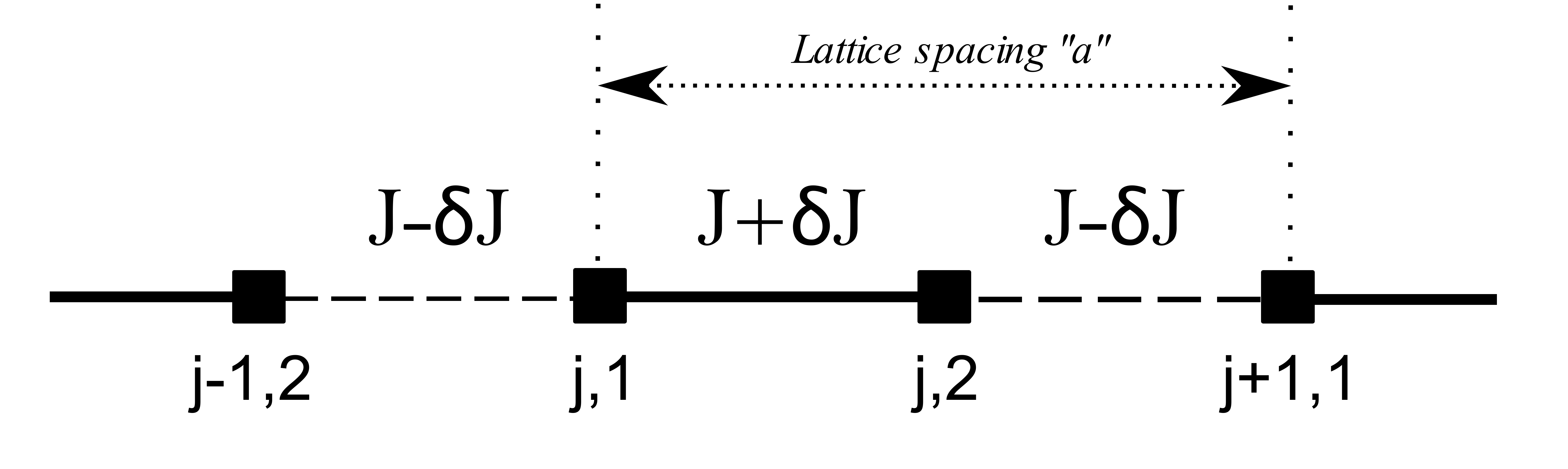}
  \caption{\label{fig:lattice_dim} Dimerized chain: to each spin 1/2 corresponds a black square, the coupling (nearest-neighbor only) is alternated with strong bonds $J_s=J+\delta J$ and weak bonds $J_w=J-\delta J$. Each spin has two indexes: the first indicating the strong bond in which it is located, the second indicating the position in the strong bond (1 for left, 2 for right). We define the lattice spacing $a$ as the distance between two unit cells, i.e.~between next-nearest neighbor sites.}
\end{figure}

In this work we consider the dimerized Heisenberg chain subjected to a magnetic field. It is described by the Hamiltonian
\begin{equation} \label{eq:dimham}
 H=\sum_j\left(J+(-1)^j\delta J\right) \mathbf{S}_j\cdot \mathbf{S}_{j+1}-h\sum_j S^z_j.
\end{equation}
$\mathbf{S}_j=\sigma_j$ is the spin (vector) at site $j$ represented by the Pauli matrices $\sigma_j$. In the following the spin operators are chosen as dimensionless and the $g$ factor, the Bohr magneton and $\hbar$  have been absorbed into the amplitudes of the magnetic field $h$, the isotropic coupling $J$ and the anisotropic coupling $\delta J$, which have the dimension of an energy. A schematic representation of this model is given in Fig.~\ref{fig:lattice_dim}.
A finite value of the anisotropic coupling leads to a dimerization of the spin coupling in the chain and we use $J_s=J+\delta J$ for the strong bonds and $J_w=J-\delta J$ for the weaker bonds. Since the unit cell has two sites we introduce the labeling of the unit cells and the two sites by the tuples $(j,r)$, where $j$ is the position of the unit cell and $r=1,2$ denotes the site within the unit cell. This labeling is represented in Fig.~\ref{fig:lattice_dim}. $L$ is the total number of sites in the chain and $a$ is the size of the unit cell.

\begin{figure}
  \centering
  \includegraphics[width=0.45\textwidth]{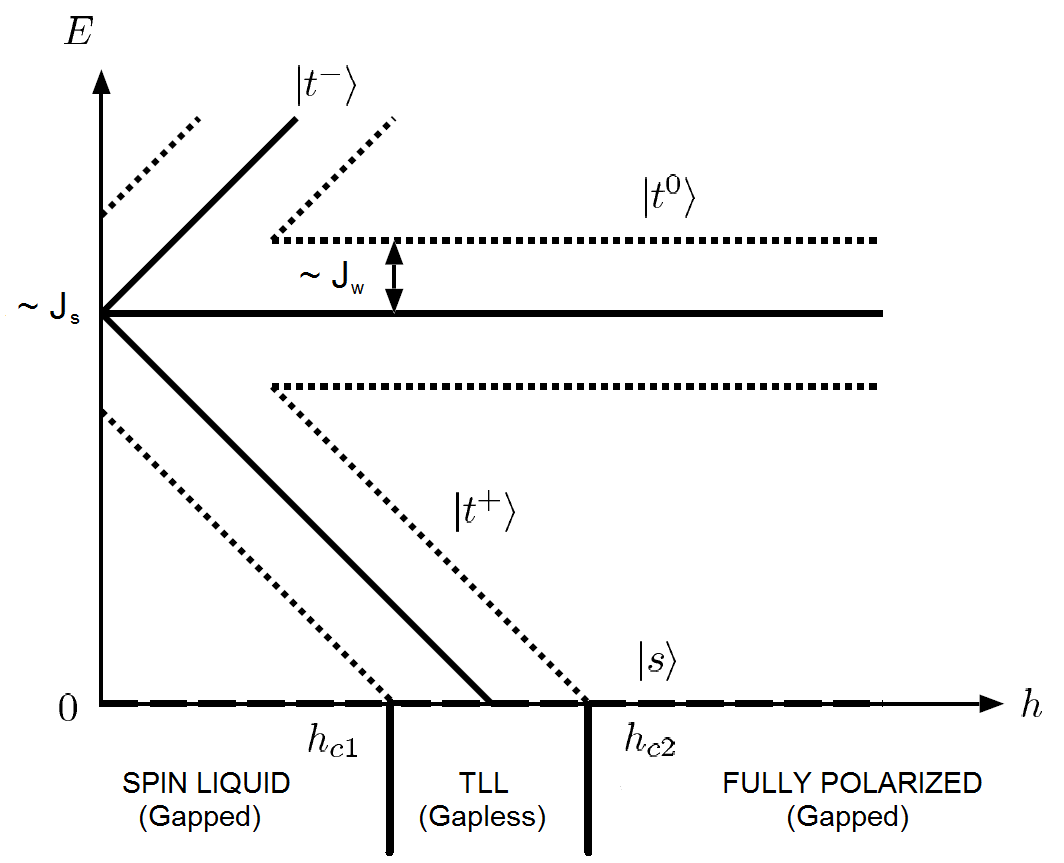}
  \caption{\label{fig:phasedi_dim} Sketch of the excitation spectrum for the dimerized chain under a magnetic field $h$. In the strong dimerization limit, i.e.~$J_w\ll J_s$, the physics of the model can be essentially restricted to that of two spins 1/2 on a strong bond. At $h=0$ there is a gap between the singlet $\left| s\right\rangle=\frac{1}{\sqrt{2}}\left(\left|\uparrow\downarrow\right\rangle-\left|\downarrow\uparrow\right\rangle\right)$, and the three triplets $\left|t^+\right\rangle=\left|\uparrow\uparrow\right\rangle$, $\left|t^0\right\rangle=\frac{1}{\sqrt{2}}\left(\left|\uparrow\downarrow\right\rangle+\left|\uparrow\downarrow\right\rangle\right)$ and $\left|t^-\right\rangle=\left|\downarrow\downarrow\right\rangle$. The field removes the degeneracy in energy of the triplets bringing down the energy of $\left| t^+\right\rangle$ (upper part) and closing the gap. Due to the weak coupling between strong bonds, a triplet can be delocalized, which results in a dispersion in energy (dotted lines). When the field is high  enough such that the lowest triplon excitation crosses the singlet excitation ($h> h_{c1}$), the triplon band begins to fill leading to a massless phase, the Tomonaga-Luttinger liquid, up to the point $h_{c2}$ where the band is totally filled and the system again forms a gap. After Fig.~2 in Ref.~\onlinecite{bouillot_dynamical_ladder}}.
\end{figure}

For $h=0$ the ground state is a non-trivial spin $0$ state and there is a gap to the first excited state which is a band of one-triplon excitations \cite{giamarchi_book_1d}. Both the ground state and the excitations can be understood in the limit of strong dimerization.
In this limit one has antiferromagnetically coupled spin dimers characterized by a strong exchange coupling $J_s$ which are themselves weakly coupled by $J_w$. The lowest energy excitations are given by a single dimer excited from spin $0$ (singlet) to spin $1$ (triplet ``+",``0'' or ``-'') and delocalized along the chain (see Fig.~\ref{fig:phasedi_dim}).
If one progressively increases the magnetic field along $z$, the gap to those excitations gets smaller and smaller up to the critical value of the field $h=h_{c1}$ where it closes. Once that this point is reached, a further increase of $h$ progressively polarizes the spins of the chain up to full polarization, occurring at $h_{c2}$. The system for $h_{c1}<h<h_{c2}$ is in a gapless phase (Tomonaga-Luttinger liquid), for $h>h_{c2}$ the phase is again gapped. For more details on dimerized chains see for instance Ref.~\onlinecite{giamarchi_book_1d} or Ref.~\onlinecite{chitra_spinchains_field}.

In this work, we concentrate on the coupling parameters pertinent for the strongly dimerized copper nitrate $\left[\text{Cu}(\text{NO}_3)_2\cdot2.5\text{D}_2\text{O}\right]$ experimentally investigated in Ref.~\onlinecite{tennant03_dimerized}. For this compound the coupling parameters have been determined to be $J\approx 3.377~k_BK$ and $\delta J\approx 1.903~k_BK$\cite{diederix_alternated_copnit,bonner_alternated_copnit}. This compound shows a relatively strong dimerization of $J_s/J_w \approx 3.58$.

\section{MPS method} \label{sec:dmrg}

In this paper we are interested in the computation of dynamic finite-temperature correlation functions of the form
\begin{equation}
\left\langle \hat{B}(t)\hat{A}\right\rangle_T = \Tr\left(\hat{\rho}_{\beta}\hat{B}(t)\hat{A}\right).
\label{eq:dmrg_corr1}
\end{equation}
Here $\hat{\rho}_\beta$ is the finite temperature density matrix, i.e.~$\hat{\rho}_{\beta} = e^{-\beta H}/Z_{\beta}$ and $\beta=1/(k_BT)$ the inverse temperature. $\hat{B}(t)=e^{iHt}\hat{B}e^{-iHt}$ is the time-evolved operator, where $H$ is the Hamiltonian of the system. We are interested in the case where $\hat{B}$ and $\hat{A}$ are spin operators ($S^+$, $S^-$, $S^z$) with the property $\hat{A}=\hat{B}^{\dagger}$.

The density matrix is encoded within a matrix product state (MPS) formalism by a pure state in an enlarged Hilbert space (real part plus auxiliary part):
\begin{equation}
 \hat{\rho}_{\beta}\longrightarrow\left|\rho_{\beta}\right\rangle\in\mathcal{H}\otimes\mathcal{H}_{\text aux}~,
\end{equation}
so that
\begin{equation}
 \Tr_{\text aux}\left|\rho_{\beta}\right\rangle\left\langle\rho_{\beta}\right|=\hat{\rho}_{\beta}.
\end{equation}
The pure state representation of the density matrix $\left|\rho_{\beta}\right\rangle$ can be determined in an MPS form by an imaginary-time evolution, starting from the initial infinite temperature state which corresponds in the pure state representation to the maximally entangled state of the form $\left|\rho_0\right\rangle\propto\sum_{\mathbf{\sigma}}\left|\mathbf{\sigma}\right\rangle\otimes\left|\mathbf{\bar{\sigma}}\right\rangle_{\text aux}$.
Here the entanglement is chosen between $\sigma$ and its opposite $\bar{\sigma}$ in order to be able to use the magnetization as a good quantum number which lightens the numerical cost of the approach.

To obtain the correlation defined in (\ref{eq:dmrg_corr1}), we make use of the computational scheme adopted in Ref.~\onlinecite{coira_nmr}, proposed originally in Ref.~\onlinecite{karrasch_tdmrg_finiteT} and analyzed in detail in Ref.~\onlinecite{barthel_tdmrg_finiteT}. The scheme can be summarized by the following expression:
\begin{equation}
 \left\langle\hat{B}(t)\hat{A}\right\rangle_T=\frac{1}{Z_{\beta}}\Tr\left([e^{-\beta H/2}]\hat{B}[e^{-iHt}\hat{A}e^{-\beta H/2}e^{iHt}]\right).
\label{eq:dmrg_corr2}
\end{equation}
Square brackets indicate which parts of this expression are approximated by an MPS. Bracketed operators are calculated with very good approximation using imaginary and/or real time evolution. Finally, the application of the local operator $\hat{B}$ has to be performed.

Typical values for bond dimensions used here range from $500$ for the lowest temperatures, up to $2000$ states for the highest ones. The minimal truncation has been chosen of the order of $10^{-20}$ for imaginary time evolution and $10^{-10}$ for real time evolution, together with a maximal truncated weight of $10^{-6}$ ($10^{-5}$ for the highest temperatures). The convergence is assured for this set of parameters (see also Appendix \ref{app:convergence}). Results shown in this work are obtained for a chain of size $L=130$, which is chosen in a way such that the perturbation created by the operator $\hat{A}$ does not reach the boundary of the system for times up to $t_{\text{max}}$ at all temperatures. The resulting finite size effects are small and therefore neglected.

\section{Dynamic correlation functions} \label{sec:procedure}

We are interested in dynamical spin-spin correlation functions at finite temperature in frequency and momentum. The described time-dependent MPS method gives access to correlations in time and space:
\begin{equation}
\left\langle S^{\lambda}_{j_2,r_2}(t_n)S^{\mu}_{j_1,r_1}(0)\right\rangle_T~=:~S^{\lambda\mu}_{r_2r_1}(d=j_2-j_1,t_n)~,
\label{eq:dmrg_corr3}
\end{equation}
where $(\lambda,\mu)$ can be $(\pm,\mp)$ or $(z,z)$. We label the sites by the unit cell index $j_i$ and the index within the unit cell $r_i$ (cf.~Fig.~\ref{fig:lattice_dim}). For each couple $(\lambda,\mu)$ there are four different correlations which we will name 11, 12, 21 and 22 following the choice for $r_2, r_1$ indices. In this work we focus on the $11$ correlations, since they are representative of the physics of the model and of the phases considered. The other correlations can be computed in the same way. The time $t_n=\delta t\cdot n$ is the time at which the second spin operator is applied. In the numerical calculation the time is discretized with a time-step $\delta t$ and $n$ being an integer. $T$ is the temperature at which the correlations are evaluated. In the numerical calculations we use a finite system and consider the correlations at $j_1=0$ and $j_2$ varying from $-\frac{\tilde{L}-1}{2}$ to $\frac{\tilde{L}-1}{2}$, where $\tilde{L}=L/2$. Given the type of correlation we are computing, translational invariance guarantees that $S^{\lambda\mu}_{r_2r_1}(d,t)=S^{\lambda\mu}_{r_2r_1}(-d,t)$. In the $(z,z)$ case we subtract $m^2$ ($m$ is the magnetization per site).

In order to obtain the correlations in the frequency domain and momentum space, two Fourier transformations are applied. In order to minimize oscillations which arise as numerical artifacts, due to finite system length and finite time $t_{\text{max}}$ reached by simulations, we overlay a gaussian filter to the time-position/space correlations before the application of the Fourier transformation following Ref.~\onlinecite{bouillot_dynamical_ladder}:
\begin{equation}
S^{\lambda\mu}_{11}(d,t_n)~\longrightarrow~S^{\lambda\mu}_{11}(d,t_n)\cdot f(d,t_n)~,
\label{eq:filters}
\end{equation}
\begin{equation}
f(d,t_n)=e^{-\left(4d/\tilde{L}\right)^2}\cdot e^{-\left(2t_n/t_{\text{max}}\right)^2}.
\label{eq:filters2}
\end{equation}

For the double Fourier transform we use the same conventions as adopted in Ref.~\onlinecite{bouillot_dynamical_ladder}:
\begin{equation}
S^{\lambda\mu}_{11}(q_k,\omega_m)\approx\delta t\sum_{n=-N_t+1}^{N_t}\sum_{d=-\frac{\tilde{L}-1}{2}}^{\frac{\tilde{L}-1}{2}}e^{i(\omega_m t_n-q_kd)}S^{\lambda\mu}_{11}(d,t_n).
\label{eq:TdF}
\end{equation}
$N_t$ is the total number of time steps made for the evolution for that specific case, $\omega_m=\frac{\pi m}{N_t\delta t}$ for $m$ integer ranging from $-N_t+1$ to $N_t$, and $q_k=\frac{2\pi k}{\tilde{L}}$ for $k=0, 1, \dots, \tilde{L}-1$. Note that the momentum is connected to unit cells containing two sites.
The negative time correlation functions are determined using the symmetries of the lattice and properties of spin operators by means of the following relation:
\begin{equation}
S^{\lambda\mu}_{11}(d,-t)=\left[S^{\lambda\mu}_{11}(d,t)\right]^*.
\end{equation}
In the following, we show the real part of these filtered correlations, $\Re \left(S^{\lambda\mu}_{11}(q,\omega)\right)$, which contains important information.

Our results for correlations can be directly connected to inelastic neutron scattering (INS) measurements as explained for instance in Ref.~\onlinecite{lovesey_neutron_scattering}. Zero temperature dynamical correlations allowed already successful comparisons with experimental results at low temperature\cite{bouillot_dynamical_ladder,schmidiger_dimpy_neutrons}. For the dimerized system, neutron scattering data has been obtained in Refs.~\onlinecite{tennant03_dimerized} and \onlinecite{stone_neutron_dimer} and with the neutron-resonance spin echo (NRSE) technique\cite{groitlhabicht2016}. By incorporating the specific details of the material (position of atoms, etc.) and summing over polarizations (since most experiments are done with unpolarized neutrons) our data can be related to the neutron absorption for a given material. These extra elements, although important if one wants to make comparison with a specific material, complicate in general the understanding. In this paper we thus directly focus on the spin-spin correlations themselves for each polarization, to analyze the effects of finite temperature.

\section{Results} \label{sec:spectra}

We present in this section our results for dynamical spin-spin correlation functions at finite temperature obtained by the time-dependent MPS method. We consider the correlations $S^{+-}_{11}$, $S^{-+}_{11}$ and $S^{zz}_{11}$. From $S^{\pm\mp}$ correlations one can easily access $S^{xx}$ and $S^{yy}$. Since we are interested in the physics of the various phases of the system (see Fig.~\ref{fig:phasedi_dim}) we consider the following magnetic fields:
\begin{itemize}
\item $h=0$ At zero magnetic field, the system is gapped. Since the system is isotropic  only the  $S^{zz}_{11}$ correlation is shown.
\item $h=2.868J_w\lesssim h_{c1}$, with $h_{c1}\sim2.976J_w$. This field lies still in the gapped regime. However, the gap is already very small since the field is very close to the critical field.
\item $h=3.716J_w$. This magnetic field puts the system well inside the massless phase. In this regime the low energy part of the spectrum is expected to be described by a Tomonaga-Luttinger liquid.
\item $h=4.674J_w > h_{c2}$, with $h_{c1}\sim4.566J_w$. In this case we are in the fully polarized regime and the system is again gapped (at zero temperature).
\end{itemize}
For each one of these regimes we examine the physics encoded in the spectra and we consider different temperatures. We choose in particular $T=0.082J_w$, $T=0.339J_w$ and $T=1.356J_w$, for which the system is still fully in the quantum regime, but temperature effects are already evident.

As we will see in detail, temperature can have different effects on the spectrum. One of the main effects of the temperature is to change, i.e. shift and smear, the dispersion relations of the excitations. We will see from the data that this broadening can be very large and in general asymmetric with respect to the zero temperature dispersion. Additionally, new excitation branches can occur due to the admixture of excited states in the finite temperature state.

On the technical side, because of the increasing complexity of the simulations for higher temperatures and for magnetic fields inside or very close to the critical region, the maximum time reached by the real-time evolution is sometimes reduced, causing a worse resolution in frequency. Nevertheless, our resolution is still sufficient to see interesting effects on the structure of the correlations.

\subsection{Isotropic gapped system $h=0$}

In absence of magnetic field the dimerized system is gapped and perfectly isotropic in the three directions $x,y,z$. Thus, we focus for $h=0$ only on the correlations along $z$-direction, i.e.~$\left\langle S^z S^z\right\rangle$. In the limit of large gap and low temperature, the application of the first operator $S^{z}_{0,1}$ creates a single triplet excitation $t_0$ on top of a background of singlets located on each strong bond. Its energy is given by $\sim J_s$. The created triplet can delocalize with time by magnetic exchange from strong bond to strong bond. The spectrum can be well approximated by the one of an isolated particle (the triplet) moving on the background of an empty lattice (the singlets) with a tight binding Hamiltonian with a hopping matrix element $-J_w/4=-(J-\delta J)/4$. More details about this result can be found in Appendix~\ref{app:strongbondpicture}. One thus expects for the spectral function a narrow cosine shape of amplitude $\sim J_w/2$.

\begin{figure}
  \centering
  \includegraphics[width=0.5\textwidth]{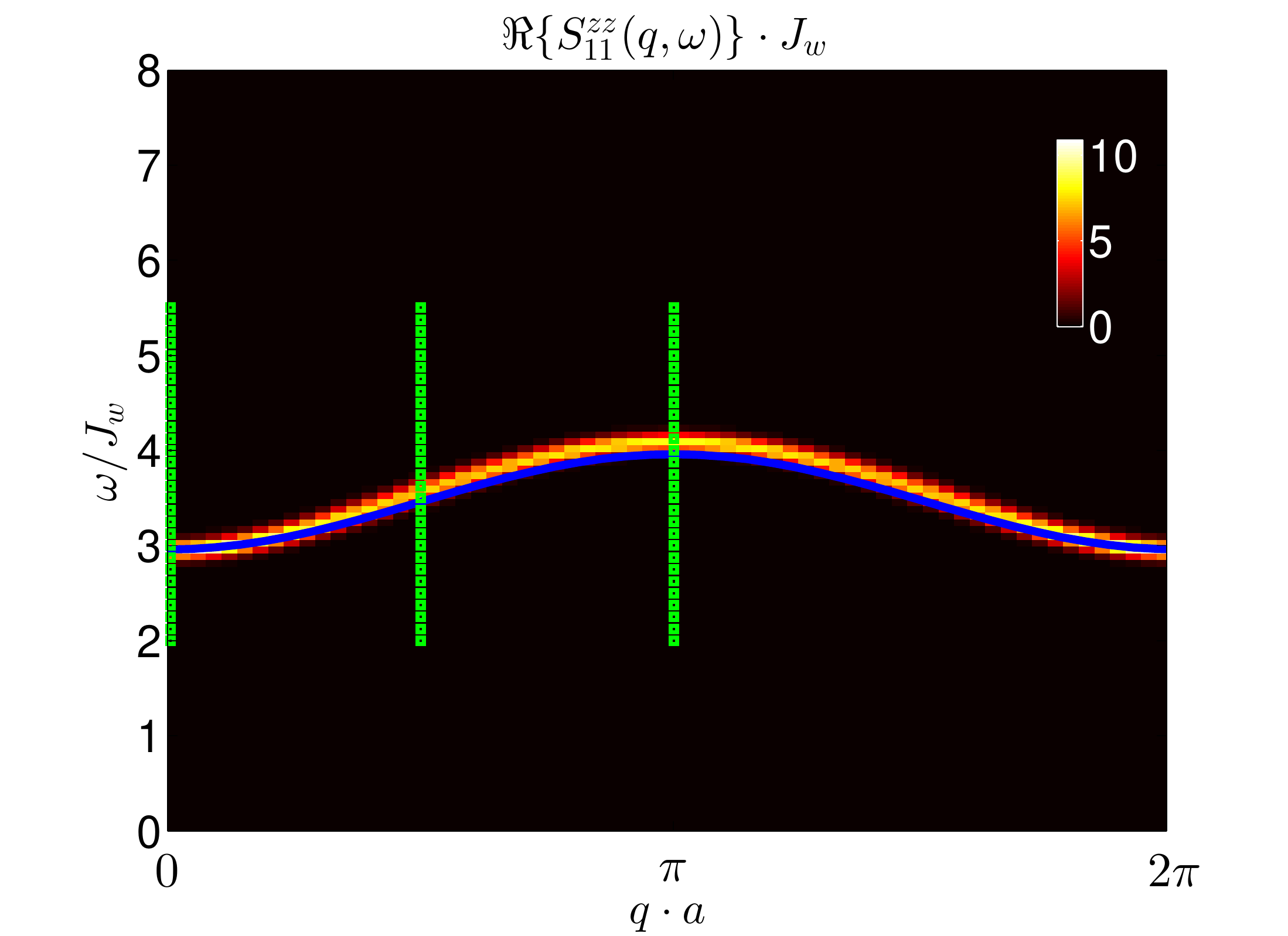}
  \caption{\label{fig:500mk_0k} Case $T=0.339J_w$, $h=0$, $\left\langle S^zS^z\right\rangle$ correlations. A cosine shaped band at finite energy shows the gapped nature of the system in this case. The blue curve represents the cosine which one gets by approximating the spectrum of the system with the one of a single triplet in a sea of singlets, hopping from one strong bond to the neighboring one with a tight binding Hamiltonian and a corresponding amplitude $J_w/2$. Green dotted lines indicate the regions where cuts have been performed to study temperature effects on the cosine band represented in Fig.~.\ref{fig:szsz0}.}
\end{figure}

\begin{figure}
  \centering
  \includegraphics[width=0.5\textwidth]{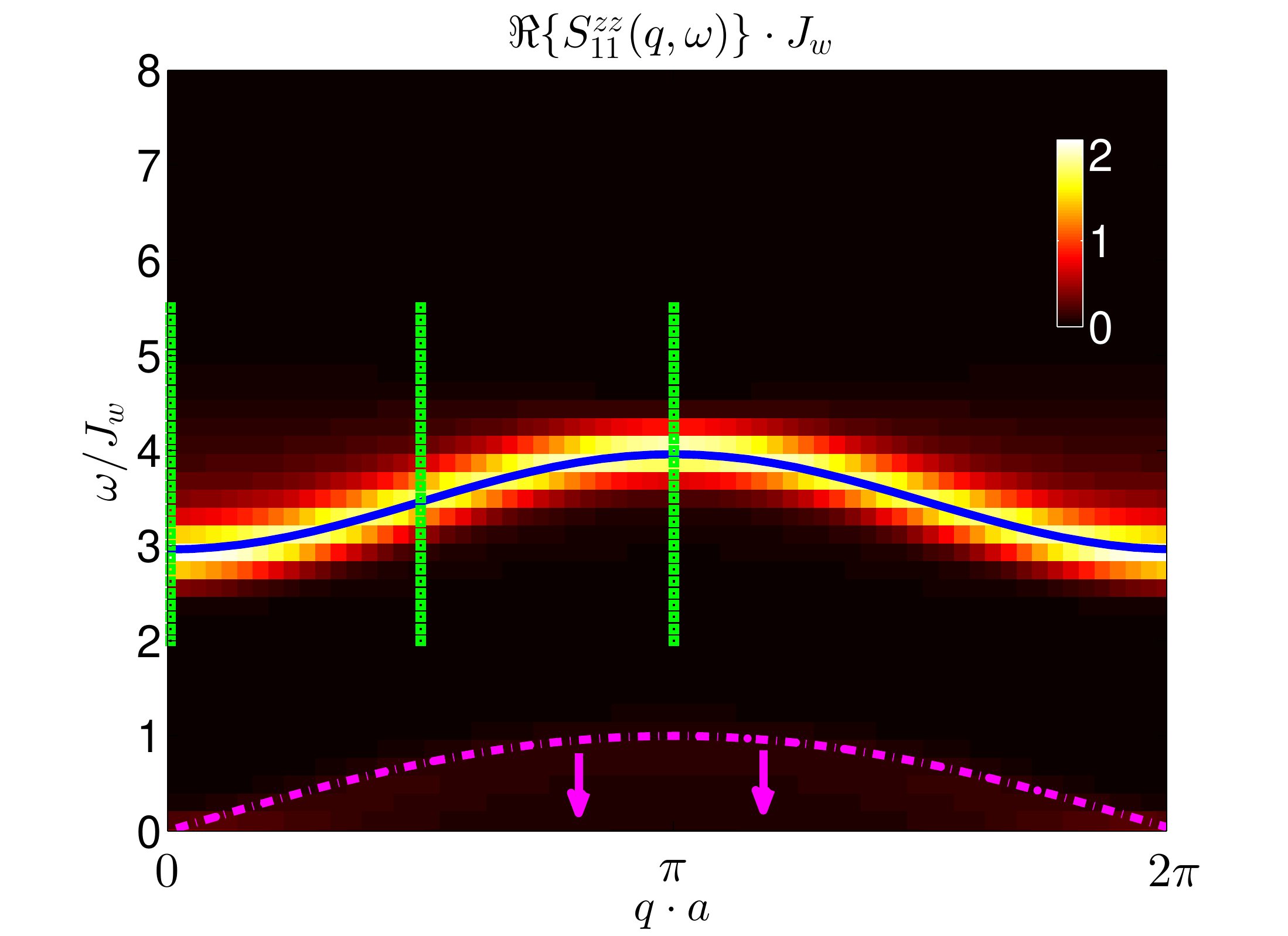}
  \caption{\label{fig:2k_0k} Case $T=1.356J_w$, $h=0$. $\left\langle S^zS^z\right\rangle$ correlations. The cosine band at finite energy is still there but with a reduction in intensity and a stronger broadening especially at $q$ close to $0$ and $2\pi/a$. New structures appear at low energy (localized below the purple dashed curve, see text and Appendix~\ref{app:lowenstru}).  Green dotted lines indicate the regions where cuts have been performed to study temperature effects on the cosine band represented in Fig.~.\ref{fig:szsz0}. }
\end{figure}

We show the $\left\langle S^z S^z\right\rangle$ correlations at $T=0.339J_w$ in Fig.~\ref{fig:500mk_0k} and at $T=1.356J_w$ in Fig.~\ref{fig:2k_0k}.

The cosine shape is well observed with an amplitude which is reasonably compatible with the value $J_w/2$. Only small modifications can be seen. The broadening in energy at each momentum value is very small at low temperature. An increase in temperature (cf.~Fig.~\ref{fig:500mk_0k} and Fig.~\ref{fig:2k_0k}) leads to mainly two different effects: additional intensity appears at low energies and a significant increase of the frequency broadening of the cosine band, connected with an important reduction of intensity. These effects had already been observed in Refs.~\onlinecite{james_thermbroad_dimer,mikeskaluckmann2006}, and stem from the fact that the initial finite temperature state is not anymore a collection of singlets on each strong bond, but can contain a sizable fraction of higher excitations (triplets ``0", ``+" or ``-") which can have finite momentum $q_i$. The possible transitions (see Appendix~\ref{app:strongbondpicture}) can take place at an energy close to zero or close to the original cosine shaped excitation band around $J_s$. By imposing conservation of energy, and taking into account the cosine dispersion which characterizes the energy $E(q)$, we find, as detailed in the Appendix~\ref{app:lowenstru}, that the weak intensity dome observed at low energy in Fig.~\ref{fig:2k_0k} is described by the relation
\begin{equation}
\frac{\omega}{J_w}=\sin\left(q_i\cdot a+\frac{q\cdot a}{2}\right)\sin\left(\frac{q\cdot a}{2}\right),
\label{eq:dome}
\end{equation}
for all possible choices of $q_i$ and for each fixed value of $q\cdot a$, being $a$ the lattice spacing. This means that (weak) intensity should be present below the purple dashed line in Fig.~\ref{fig:2k_0k} which is indeed what it is observed.

Let us now examine in detail how the temperature can affect the cosine dispersion of triplet excitations.
In order to highlight the thermal effects found in our numerics, we make cuts at fixed values of momenta $q\cdot a=0, \pi/2, \pi$ on the intensity plots of $\left\langle S^z S^z\right\rangle$ correlations, around the cosine band, at various temperatures, ranging from $T=0.082J_w$ up to $T=6.78J_w$. These cuts correspond for example to the green dotted lines in figures \ref{fig:500mk_0k} and \ref{fig:2k_0k}.
\begin{figure*}
  \centering
  \includegraphics[width=0.925\textwidth]{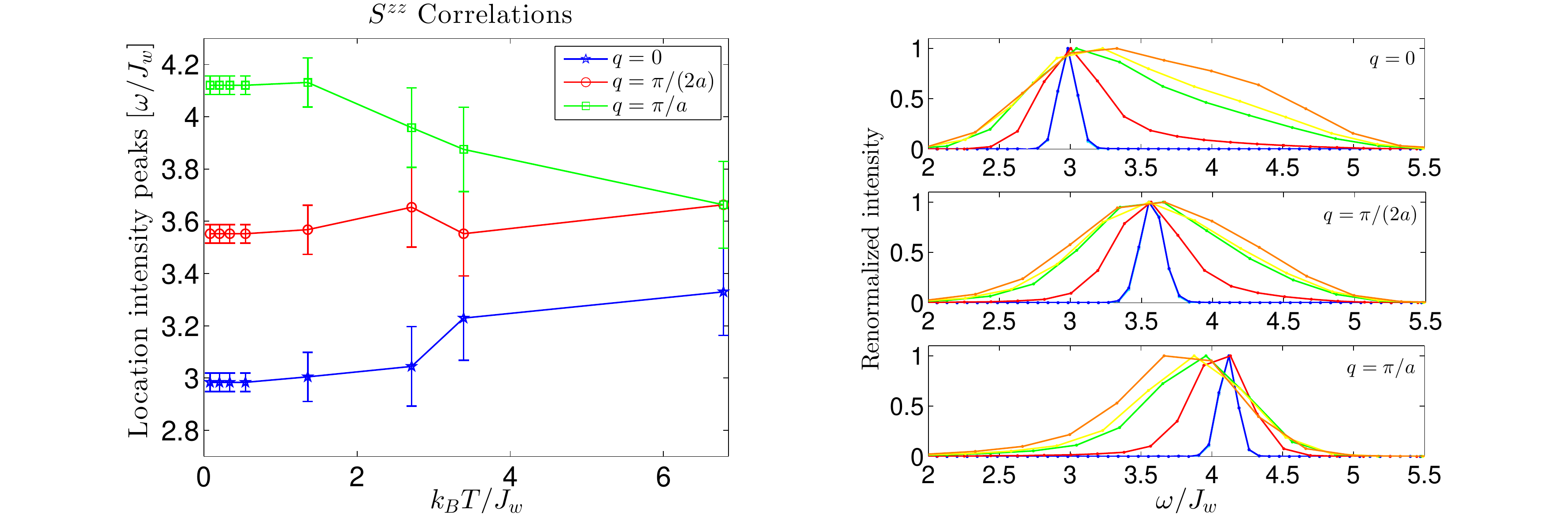}
  \caption{\label{fig:szsz0} Position of intensity maxima at $q=0, \frac{\pi}{2}, \pi$ as a function of $k_BT/J_w$ for $\left\langle S^zS^z\right\rangle$ correlations in the case $h=0$ (left plot).
On the right, cuts at fixed momentum as a function of energy, as obtained {\it i.e.} from Fig.~\ref{fig:500mk_0k} or Fig.~\ref{fig:2k_0k}. All intensities have been renormalized in order to have 1 as a maximum. A different color corresponds to a different temperature with black $\leftrightarrow0.082J_w$, purple $\leftrightarrow0.204J_w$, light blue $\leftrightarrow0.339J_w$, blue $\leftrightarrow0.543J_w$, red $\leftrightarrow1.356J_w$, green $\leftrightarrow2.712J_w$, yellow $\leftrightarrow3.39J_w$, orange $\leftrightarrow6.78J_w$. Curves from $T=0.082J_w$ to $T=0.543J_w$ are all on top of each other (see blue curve).}
\end{figure*}
We report our results in Fig.~\ref{fig:szsz0}, which is organized in two parts. On the left, for each of the three values of momentum, we plot the positions of the maxima in intensity of the cosine band at different temperatures. The error bars are given by the available resolution in energy (which is related to the $t_{max}$ reached in the simulations). In the three plots on the right we report the cuts made for the three values of momenta, respectively, as a function of energy for different temperatures. The intensities have all been renormalized to one in order to emphasize the broadening. As the temperature increases a strong \emph{asymmetric} broadening of the spectra appears and due to this asymmetric broadening the maxima of the curves shift slightly in energy.

The broadening is caused by the presence of different triplet excitations in the thermal state. Due to this, more transitions at slightly different energies are possible which leads to the broadening of the spectrum. The asymmetry in the broadening is coming from the energy-momentum constraints.  Compared to the results in Ref.~\onlinecite{james_thermbroad_dimer} the main findings of an asymmetric broadening are similar. A direct comparison is difficult due to the different choice of parameters. This asymmetric broadening has been investigated also in Refs.\onlinecite{fauseweh_diagram1,fauseweh_diagram2,fauseweh_dmrg_exp} using a diagrammatic method, and shown to be a universal feature in many quantum magnets.

\subsection{Gapped regime at finite $h$}

\begin{figure*}
  \centering
  \includegraphics[width=0.925\textwidth]{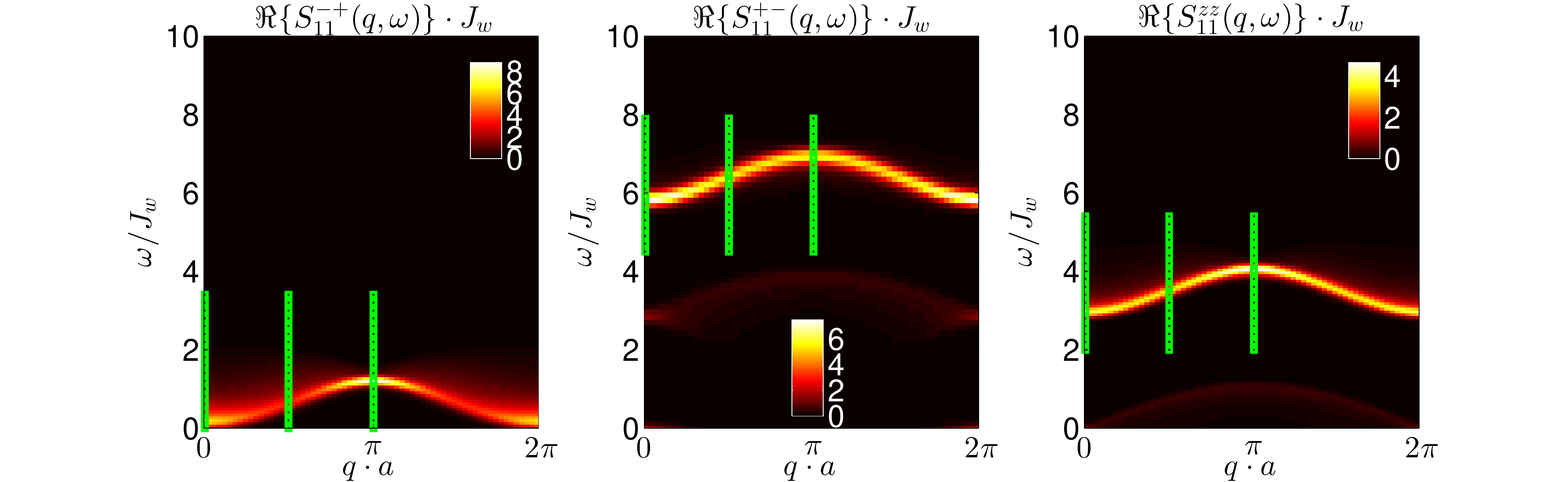}
  \caption{\label{fig:500mk_4_227k} Case $T=0.339J_w$, $h=2.868J_w$. $\left\langle S^-S^+\right\rangle$ (left), $\left\langle S^+S^-\right\rangle$ (center) and $\left\langle S^zS^z\right\rangle$ (right) correlations. The gap of the system is almost closed by the magnetic field. Cosine bands displace depending on the correlation considered and temperature effects (broadening, lower energy structures) can be already seen. Green dotted lines indicate the regions where cuts have been performed to study temperature effects on the cosine band. }
\end{figure*}

\begin{figure*}
  \centering
  \includegraphics[width=0.925\textwidth]{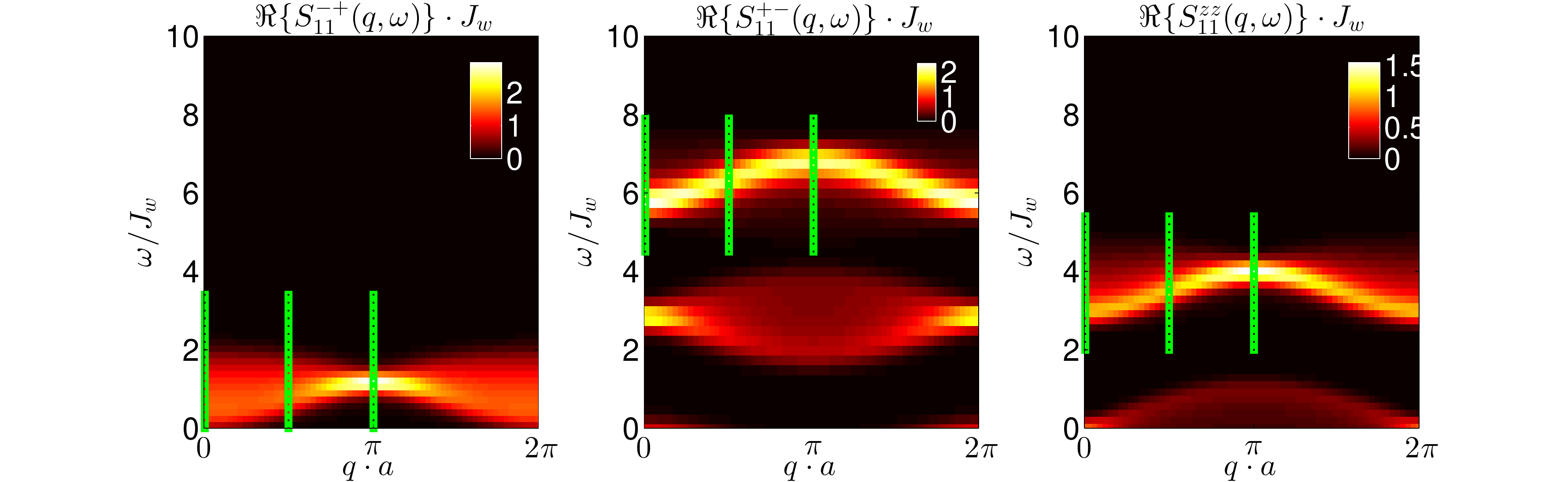}
  \caption{\label{fig:2k_4_227k} Case $T=1.356J_w$, $h=2.868J_w$. $\left\langle S^-S^+\right\rangle$ (left), $\left\langle S^+S^-\right\rangle$ (center) and $\left\langle S^zS^z\right\rangle$ (right) correlations. Temperature effects are enhanced, low energy features look more pronounced. For $\left\langle S^-S^+\right\rangle$ correlations the cosine band is substantially lost especially close to $q=0$ and $q=2\pi/a$. Green dotted lines indicate the regions where cuts have been performed to study temperature effects on the cosine band. }
\end{figure*}

At $h=2.868J_w<h_{c1}$ the gap between the ground state and the first excitation band is almost closed and the isotropy present at $h=0$ is lifted.
Thus, the three correlations $zz$, $+-$, and $-+$ behave differently as shown in Fig.~\ref{fig:500mk_4_227k} and
Fig.~\ref{fig:2k_4_227k} for the two different temperatures.
In particular, the cosine shaped spectrum is shifted strongly in energy between the three correlations.
This can be qualitatively understood in the single strong dimer picture. The application of $S^+$ to the first site of the dimer induces a transition $\left|s\right\rangle\rightarrow\left|t^+\right\rangle$. The excitation energy of the $\left|t^+\right\rangle$ triplet was brought down by the magnetic field and is already very small for the considered magnetic field.
This explains why the cosine band for $\left\langle S^-S^+\right\rangle$ correlations drops to almost zero energy in the left plot of Fig.~\ref{fig:500mk_4_227k}. The weight within the cosine band has already changed and maximal weight is found around $q\approx 0$ and $q\approx a\pi$. This is a precursor of the Tomonaga-Luttinger liquid spectrum occurring above $h_{c1}$, where gapless modes are present at $q=0$ with a  significan weight and the remaining weight of the spectrum decreases. At finite temperature a broadening mainly around these points of high intensity can be observed.  Additionally, one could expect a band arising at the energy of the $\ket{t^-}$ excitation which, however, at the shown temperatures has a negligible weight. Following the same line of reasoning, the application of $S^-$ to the first site of the dimer induces a transition $\left|s\right\rangle\rightarrow\left|t^-\right\rangle$ at energy $E_-$ which rises as $h$ grows. This explains why the cosine band for $\left\langle S^+S^-\right\rangle$ now climbs to higher energy in the central plot of Fig.~\ref{fig:500mk_4_227k}. At a finite initial temperature, a sizable fraction of triplet states $\ket{t^+}$ is admixed which can be transformed to a superposition of $\ket{s}$ and $\ket{t^0}$ which leads to the additional feature at energies around $E_0\propto J_s$.
Finally, the application of $S^z$ to the first site of the dimer induces a transition $\left|s\right\rangle\rightarrow\left|t^0\right\rangle$ which is not sensitive in terms of energy to the application of $h$. This explains why the cosine band for $\left\langle S^zS^z\right\rangle$ remains at the same energy $\approx E_0 \propto J_s$ in the right plot of Fig.~\ref{fig:500mk_4_227k}. In this correlation, the admixture of the triplet states in the finite temperature state leads to an energy band arising close to zero energy, to which the transitions of $\ket{t^+} \rightarrow \ket{t^+}$ and $\ket{t^0} \rightarrow \ket{s}$ can contribute.

The effect of the finite temperature not only shows up in the additionally allowed excitations, but it also has an effect on the form of the excitation bands which initially are cosine shaped. Already at $T=0.339J_w$ some temperature effects are clearly seen in the broadening of the cosine bands. At a higher temperature (Fig.~\ref{fig:2k_4_227k}) these effects are enhanced, and the maxima of intensity of the signal are much lower and redistributed.

\begin{figure*}
  \centering
  \includegraphics[width=0.925\textwidth]{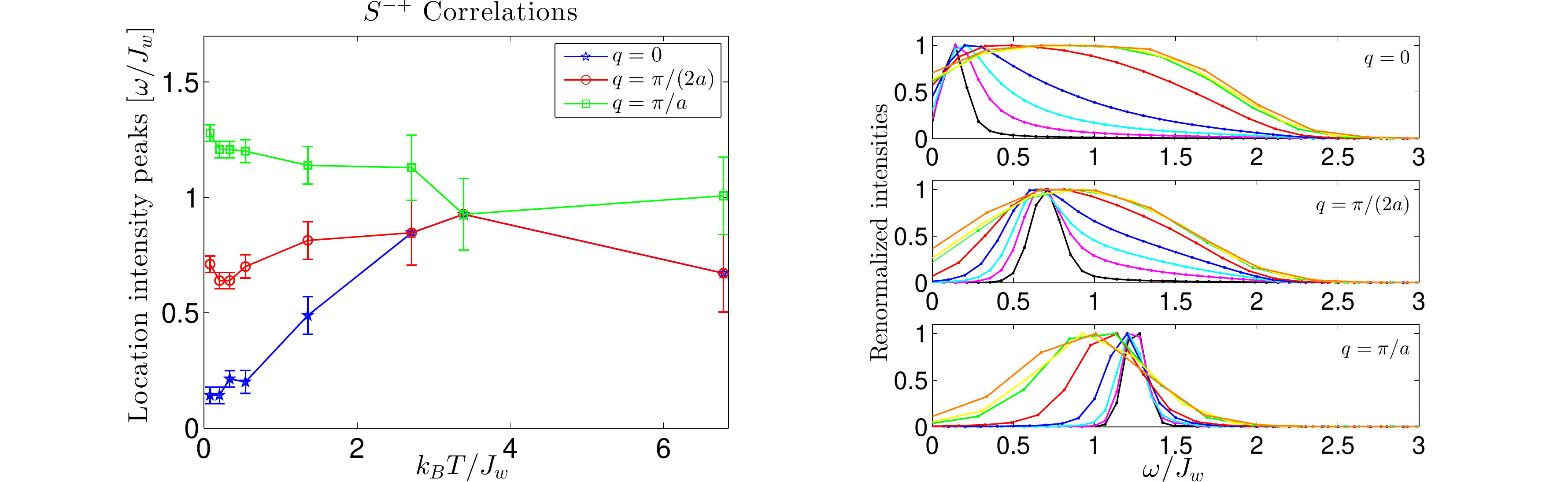}
  \caption{\label{fig:smsp4_227} Position of intensity maxima at $q=0, \frac{\pi}{2}, \pi$ as a function of $k_BT/J_w$ for $\left\langle S^-S^+\right\rangle$ correlations in the case $h=2.868J_w$ (left plot). The squeezing of the cosine band is clearly seen and a non-trivial behavior is observed at low T. On the right, cuts at fixed momentum as a function of energy, as obtained {\it i.e.} from Fig.~\ref{fig:500mk_4_227k} or Fig.~\ref{fig:2k_4_227k}. All intensities have been renormalized in order to have 1 as a maximum. A different color corresponds to a different temperature as in Fig.~\ref{fig:szsz0}.}
\end{figure*}

\begin{figure*}
  \centering
  \includegraphics[width=0.925\textwidth]{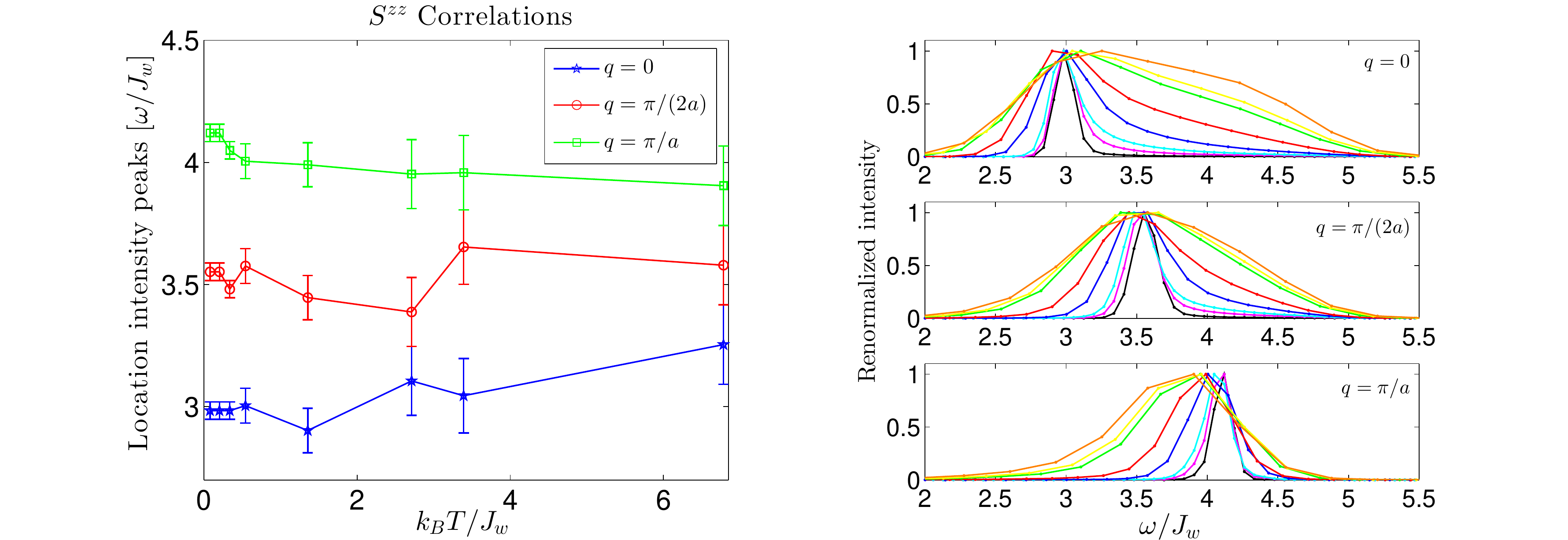}
  \caption{\label{fig:szsz4_227} Position of intensity maxima at $q=0, \frac{\pi}{2}, \pi$ as a function of $k_BT/J_w$ for $\left\langle S^zS^z\right\rangle$ correlations in the case $h=2.868J_w$ (left plot). The squeezing of the cosine band is clearly seen and a non-trivial behavior is observed at low T. On the right, cuts at fixed momentum as a function of energy, as obtained {\it i.e.} from Fig.~\ref{fig:500mk_4_227k} or Fig.~\ref{fig:2k_4_227k}. All intensities have been renormalized in order to have 1 as a maximum. A different color corresponds to a different temperature as in Fig.~\ref{fig:szsz0}.}.
\end{figure*}

In order to analyse in more detail the change of the dominant cosine shaped dispersion we make cuts at fixed values of momenta (see green dotted lines in Figs.~\ref{fig:500mk_4_227k} and \ref{fig:2k_4_227k}). Results are shown in Figs.~\ref{fig:smsp4_227} $\rightarrow$ \ref{fig:szsz4_227}.
\begin{figure*}
  \centering
  \includegraphics[width=0.925\textwidth]{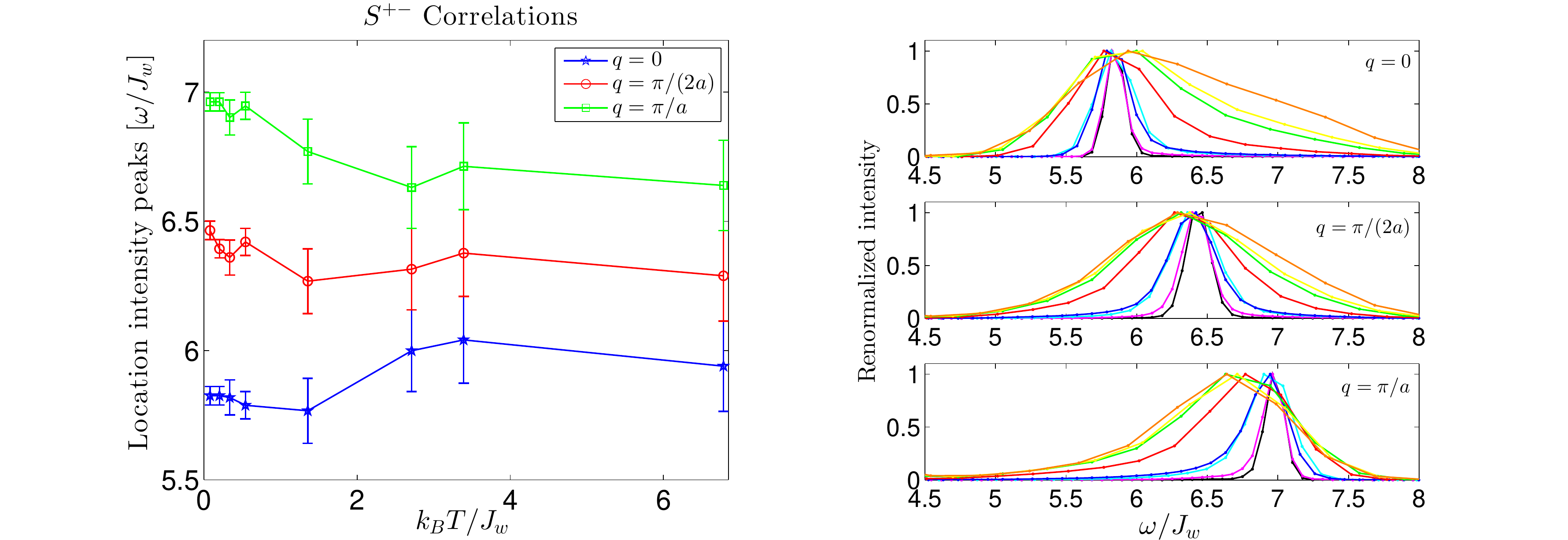}
  \caption{\label{fig:spsm4_227} Position of intensity maxima at $q=0, \frac{\pi}{2}, \pi$ as a function of $k_BT/J_w$ for $\left\langle S^+S^-\right\rangle$ correlations in the case $h=2.868J_w$ (left plot). The squeezing of the cosine band is clearly seen and a non-trivial behavior is observed at low T. On the right, cuts at fixed momentum as a function of energy, as obtained {\it i.e.} from Fig.~\ref{fig:500mk_4_227k} or Fig.~\ref{fig:2k_4_227k}. All intensities have been renormalized in order to have 1 as a maximum. A different color corresponds to a different temperature as in Fig.~\ref{fig:szsz0}.}
\end{figure*}
Again, the error bars on the left plot are given by the available resolution in frequency/energy.
The intensities have all been renormalized to one in order to emphasize the features of the broadening effect.
The asymmetric broadening is clearly visible in particular around $q=0$. By this asymmetric broadening the maximum of the weight shifts to different frequencies (as seen in the left plot).

Note that, surprisingly, this is the case even when the temperature is relatively small compared to the energy of the corresponding excitation. This is for example clear in Fig.~\ref{fig:szsz4_227} in which a relatively moderate temperature $T=1.356J_w$ is not only modifying very seriously parts of the spectra at energies of the order of $\omega=4J_w$ (the green curves in the right plots of Fig.~\ref{fig:szsz4_227}) but is also able to produce quite significant broadening of the order of $3.5J_w$. One should thus take with a grain of salt the standard qualitative estimate that the effects of the temperature are negligible when the energy of the excitation $E$ is such that $E \gg T$.

\subsection{Tomonaga-Luttinger liquid regime}

\begin{figure*}
  \centering
  \includegraphics[width=0.925\textwidth]{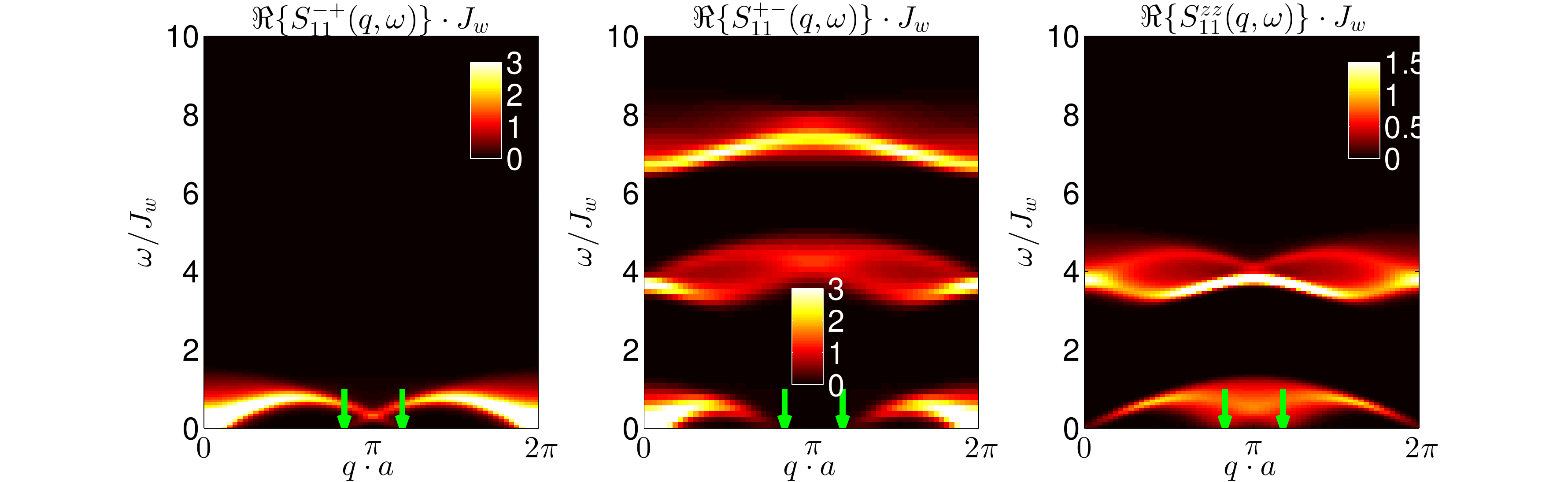}
  \caption{\label{fig:120mk_5_479k} Case $T=0.082J_w$, $h=3.716J_w$. $\left\langle S^-S^+\right\rangle$ (left), $\left\langle S^+S^-\right\rangle$ (center) and $\left\langle S^zS^z\right\rangle$ (right) correlations. Green arrows indicate the positions of the zero energy points according to the low energy approximated description (mapping to an homogeneous spin chain). The agreement between the numerical results and the theoretical expectations is excellent.}
\end{figure*}

\begin{figure*}
  \centering
  \includegraphics[width=0.925\textwidth]{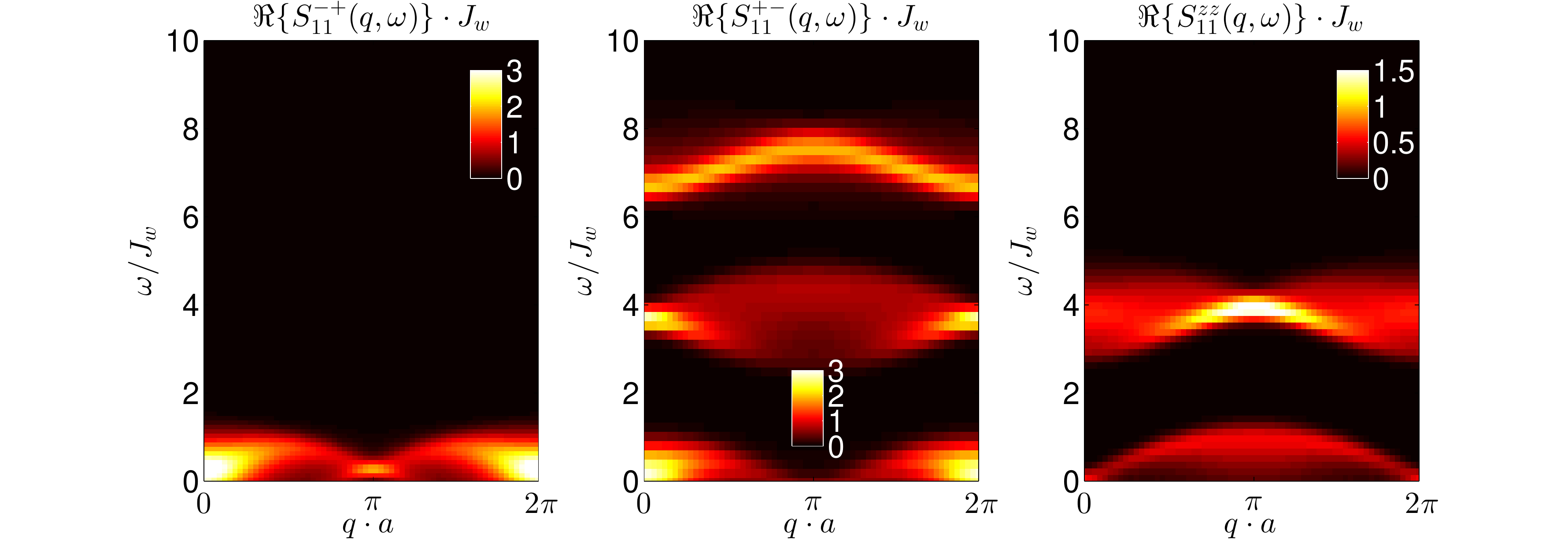}
  \caption{\label{fig:500mk_5_479k} Case $T=0.339J_w$, $h=3.716J_w$. $\left\langle S^-S^+\right\rangle$ (left), $\left\langle S^+S^-\right\rangle$ (center) and $\left\langle S^zS^z\right\rangle$ (right) correlations. The temperature starts to play a big role with respect to the previous picture, deforming the structures and redistributing the intensities.}
\end{figure*}

At $h=3.716J_w$, deep in the TLL region and at finite magnetization, structures become much more complex. At low temperature and very low energy the structures can be understood if we consider a mapping to an homogeneous spin chain (single strong dimer $\leftrightarrow$ spin). We know from Ref.~\onlinecite{giamarchi_book_1d,chitra_spinchains_field} that at finite magnetization and at low enough temperature, for an homogeneous spin-1/2 chain, $\left\langle S^zS^z\right\rangle$ correlations have low energy modes at $q=0, 2\pi$ and at $q=\pi(1\pm2m)$, while XY correlations develop incommensurability at $q=2\pi m$ and $q=2\pi(1-m)$, while the $q=\pi$ point stays commensurate. Here with $m$ we indicate the magnetization per site of the homogeneous chain. Taking into account the proper mapping between the dimerized chain and the homogeneous one, and keeping in mind that we are computing correlations of ``11'' species (see above), we find in our framework that for all the three correlations $\left\langle S^-S^+\right\rangle$, $\left\langle S^+S^-\right\rangle$ and $\left\langle S^zS^z\right\rangle$ zero energy points should sit on momenta $q\cdot a=0,~4\pi m_d,~2\pi(1-2m_d),~2\pi$. Here $m_d$ is the magnetization per site of the dimerized chain. The theoretical expectation are in very good agreement with the numerical results as shown in Fig.~\ref{fig:120mk_5_479k}.
In Fig.~\ref{fig:500mk_5_479k}
we see how the thermal effects affect these structures. The signal becomes more diffuse and the intensity is redistributed.
Whereas for the gapped phase only slight temperature effects were seen for $T=0.339J_w$, in the TLL already drastic changes are seen for this temperature.

\subsection{High field gapped phase}

\begin{figure*}
  \centering
  \includegraphics[width=0.925\textwidth]{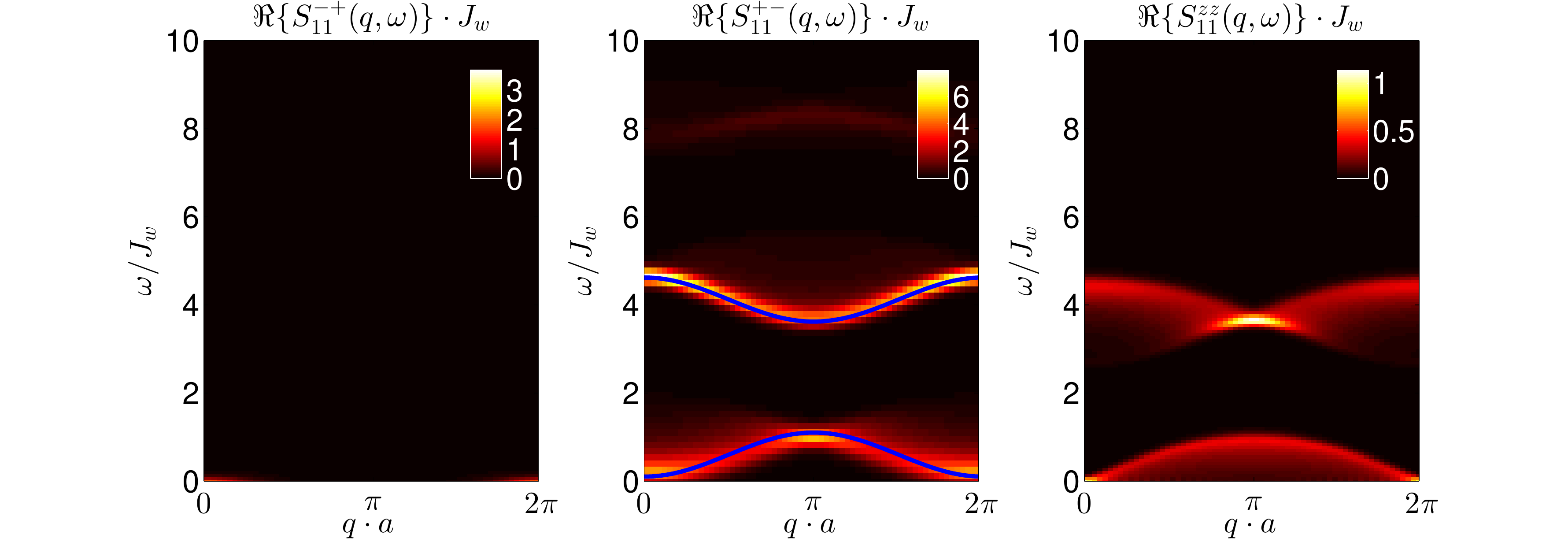}
  \caption{\label{fig:500mk_6_888k} Case $T=0.339J_w$, $h=4.674J_w$. $\left\langle S^-S^+\right\rangle$ (left), $\left\langle S^+S^-\right\rangle$ (center) and $\left\langle S^zS^z\right\rangle$ (right) correlations. At relatively low temperature $\left\langle S^-S^+\right\rangle$ correlations are zero because all spins are up. In the central plot two cosines both with amplitude $J_w/2$ are superposed to the structures we associate to the $\ket{s}$ (lower cosine) and to the $\ket{t^0}$ (upper cosine) excitations, according to the single strong dimer treatment.}
\end{figure*}

\begin{figure*}
  \centering
  \includegraphics[width=0.925\textwidth]{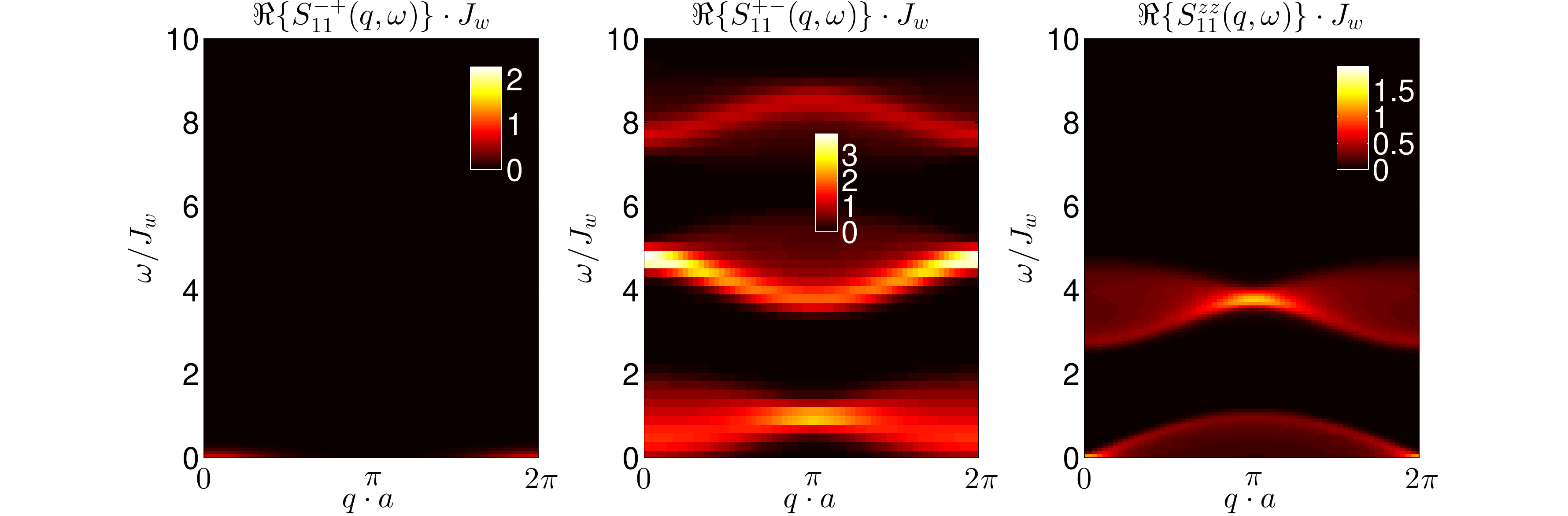}
  \caption{\label{fig:2k_6_888k} Case $T=1.356J_w$, $h=4.674J_w$. $\left\langle S^-S^+\right\rangle$ (left), $\left\langle S^+S^-\right\rangle$ (center) and $\left\langle S^zS^z\right\rangle$ (right) correlations. Some signal is now present in $\left\langle S^-S^+\right\rangle$ correlations at low energy and $q\sim0$ or $2\pi/a$ because thermal fluctuations can make some spin flips in the system. The other two correlations get broadened and there is a different redistribution of relative intensity.}
\end{figure*}

At $h=4.674J_w$, the system becomes gapped and fully polarized, i.e.~only $\ket{t^+}$ triplets are present in the ground state.
Therefore,  we expect $\left\langle S^-S^+\right\rangle$ and $\left\langle S^zS^z\right\rangle$
correlations to have zero weight at low temperature as seen in Fig.~\ref{fig:500mk_6_888k}. However, at finite temperature also singlets or other triplets can be present in the thermal state and we expect excitations at low energy to be possible, e.g. by the transitions $S^z \ket{s}\rightarrow\ket{t_0}$ at energies $\omega \approx J_s$, or $S^z \ket{t_0}\rightarrow\ket{s}$ and $S^+\ket{s}\rightarrow \ket{t^+}$ both at energies $\omega \approx 0$. As seen in Fig.~\ref{fig:2k_6_888k} these arise and have already a considerable weight at low temperatures. This is because the gap at this specific value of the field is pretty small and allows excitations already at low temperatures.

In contrast in the correlations $\left\langle S^+S^-\right\rangle$ already at zero temperature a pronounced signal is expected which can again be understood qualitatively in the single strong dimer picture: at zero temperature the system is made by a collection of $\ket{t^+}$ on each strong bond, the application of $S^-$ on a site induces a transition on that strong bond to a state which is a combination of a singlet $\ket{s}$ and of a triplet $\ket{t^0}$. Approximately, cosine shaped bands are representative of these excitations: the lower one for the singlet, which shows a very small gap compatible with the value $h-h_{c2}$, the upper one for the $\ket{t^0}$. More details on this result can be found in Appendix~\ref{app:strongbondpicture}.
At finite temperature, thermal excitations can be generated in the initial state. The applications of $S^-$ to one of these excitations (for example a singlet) can create a high-energy $\ket{t^-}$ which propagates on the chain and generates the observed faint structure at high energy.

\section{Conclusion and outlook} \label{sec:conclusions}

In this paper we have used time dependent matrix product state techniques to compute the correlation functions of a dimerized system of spins $1/2$ as a function of frequency, momentum and temperature. We have analyzed the magnetic field and temperature effects on such correlations. Our system at small magnetic field is in a gapped regime, it becomes a Tomonaga-Luttinger liquid when $h$ exceeds the first critical value $h_{c1}$, and then it becomes again gapped for $h>h_{c2}$ (fully polarized regime). For each of these cases we offered a qualitative interpretation of most structures seen using the strong coupling picture. We have investigated how the temperature broadens the spectrum and redistributes intensity among the excitations. In addition we have studied the effects of the temperature on the dispersion of a single triplet excitation in the gapped regime. We could compute quantitatively such a broadening and showed that contrarily to the results of some approximate techniques such as a bond operator technique, a strong asymmetric broadening exists at finite temperature. This effect is largely dominant over other effects such as band narrowing. In addition to the new structures, this broadening is the main temperature effect and makes it impossible to represent excitations as well defined $\delta$ function excitations even at moderate temperatures compared with the spin exchange.

Our study has been carried out for an Hamiltonian which explicitly breaks translational invariance, but similar dimerizations can occur also in models such as the $J_1-J_2$ model (with a next nearest neighbor interaction) above a certain threshold. There are however important differences at the level of the excitations. Indeed for the dimerized model the excitation is a triplet on the rung, while for spontaneously dimerized system the excitation can be a spin 1/2 followed by a shift by one lattice spacing of the dimerized bonds. In the bosonization language the operator appearing in the Hamiltonian of our model is a $\sin(2\phi)$ operator (see e.g. Ref.\onlinecite{giamarchi_book_1d} of the paper), while for the $J_1-J_2$ model this would be a $\cos(4\phi)$. The corresponding topological excitation thus carries a spin $S=1$ for the dimerized model and $S=1/2$ for the spontaneously dimerized one. So although the two models share common elements (presence of a gap, ground state made of dimers etc.) they also show important differences in their excitations. It would be therefore interesting to analyze the temperature effects on the $J_1-J_2$ model for the later.

Our calculations are potentially directly relevant to neutron scattering experiments that have been done on compounds such as the copper nitrate  $\left[\text{Cu}(\text{NO}_3)_2\cdot2.5\text{D}_2\text{O}\right]$ investigated in Refs.~\onlinecite{tennant03_dimerized} and \onlinecite{stone_neutron_dimer}. In order to go from the results of the present paper to a comparison with the experimental results several additional ingredients must be taken into account. The various correlations are mixed due to the use of unpolarized neutrons and the Fourier transform for the neutrons should be made properly, taking into account the positions of the spins in the real material (carried on $\text{Cu}$ sites that are not simply arranged in a straight line). None of these complications is however a major one and we expect most of the effects detected here directly on the individual correlations, such as the broadening by temperature, to survive. A direct comparison would thus be very interesting.

On a more general level it proves that accessing the thermal effects for realistic parameters and models that go beyond the simple case of a spin chain is now possible. The accuracy of the  numerical calculation is, like for the case of zero temperature\cite{bouillot_dynamical_ladder}, sufficient for an excellent description of neutron spectra with the usual typical resolution of the order of the meV. Thus for systems such as $\left[\text{Cu}(\text{NO}_3)_2\cdot2.5\text{D}_2\text{O}\right]$ or the spin ladders for which several neutron studies are already present in the literature, a direct comparison between the experiments and our computed spectra should be possible. This will not be the case for compounds in which the interchain or interladder coupling is stronger and the system is not in the pure one dimensional limit anymore. For such systems new methods to deal with the interchain coupling must be developed.

\begin{acknowledgements}
We acknowledge fruitful discussions with J.-S. Caux and P. Bouillot. This work was supported in part by the Swiss National Science Foundation under Division II, the DFG (TR 185 project B4, SFB 1238 project C05, and Einzelantrag) and the ERC (Grant No.648166) (C.K.).
\end{acknowledgements}

\appendix

\section{Convergence checks}
\label{app:convergence}
\begin{figure*}
  \centering
  \includegraphics[width=1\textwidth]{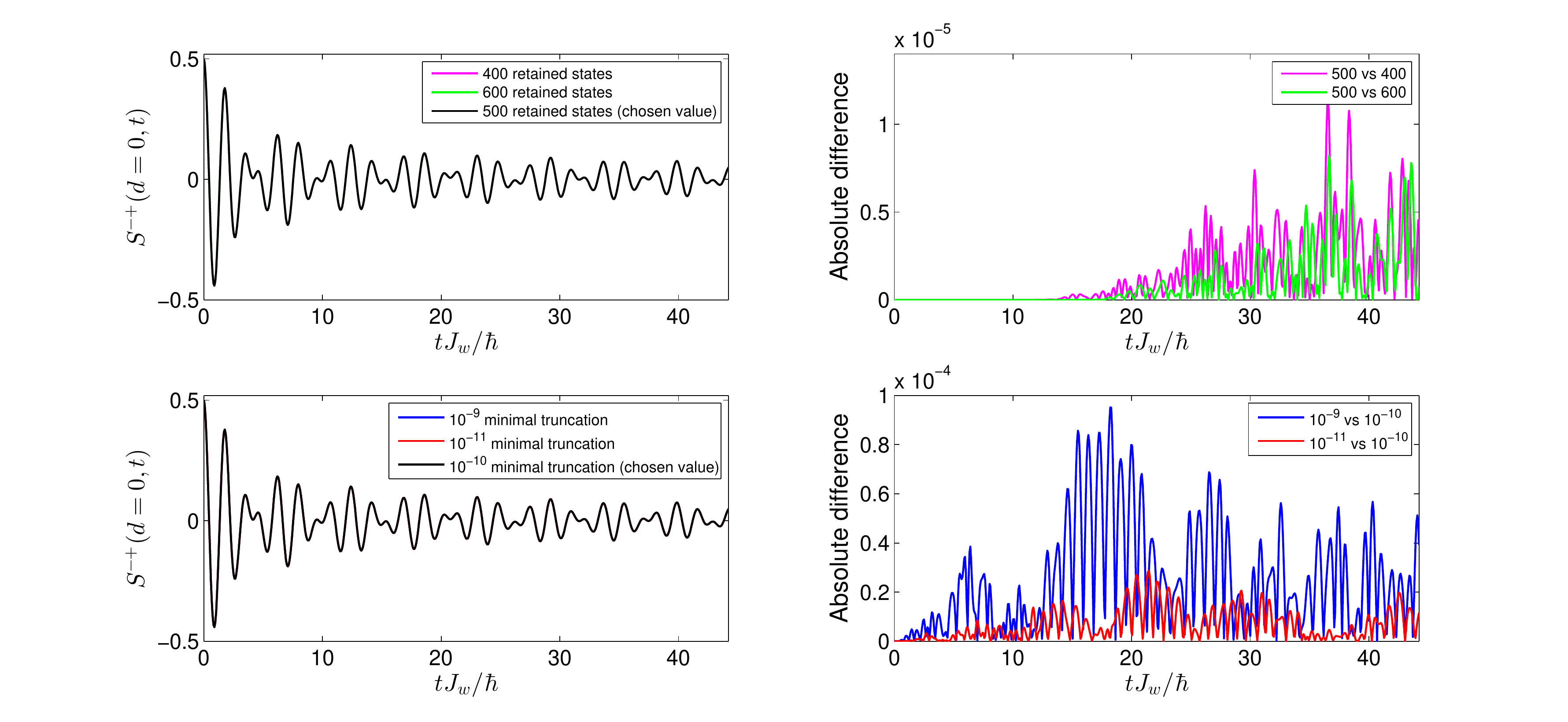}
  \caption{\label{fig:0_0k} Convergence checks for $h=0$ (gapped regime) at $T=0.339J_w$. $\left\langle S^-S^+\right\rangle(d=0,t)$ correlations (left) and the corresponding differences to the chosen value (right) plotted as a function of time, for three different values of retained states (top) and minimal truncation (bottom).}
\end{figure*}

\begin{figure*}
  \centering
  \includegraphics[width=1\textwidth]{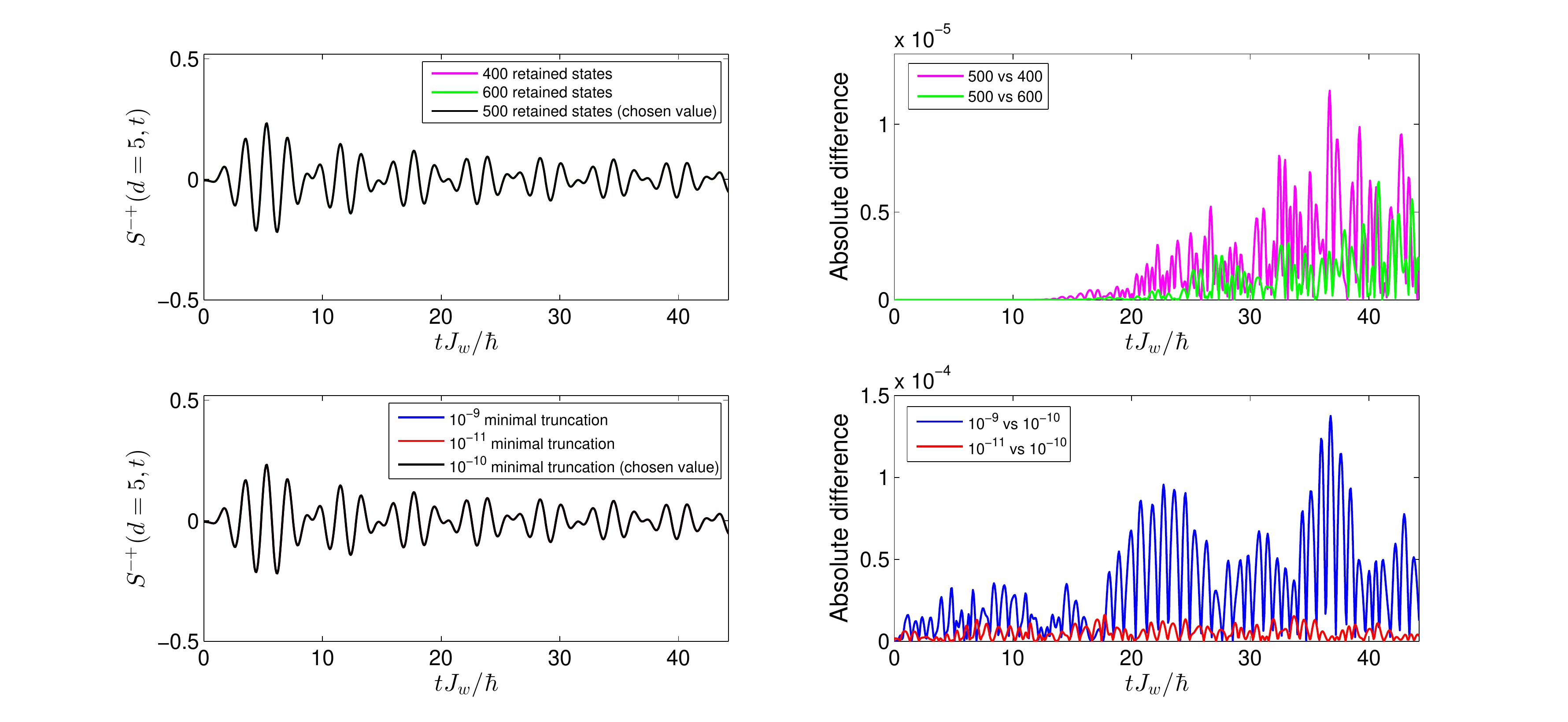}
  \caption{\label{fig:5_0k} Convergence checks for $h=0$ (gapped regime) at $T=0.339J_w$. $\left\langle S^-S^+\right\rangle(d=5,t)$ correlations plotted as a function of time (left), for three different values of retained states (top) and minimal truncation (bottom). On the right, plot of the absolute difference between the selected value for that specific parameter.}
\end{figure*}

\begin{figure*}
  \centering
  \includegraphics[width=1\textwidth]{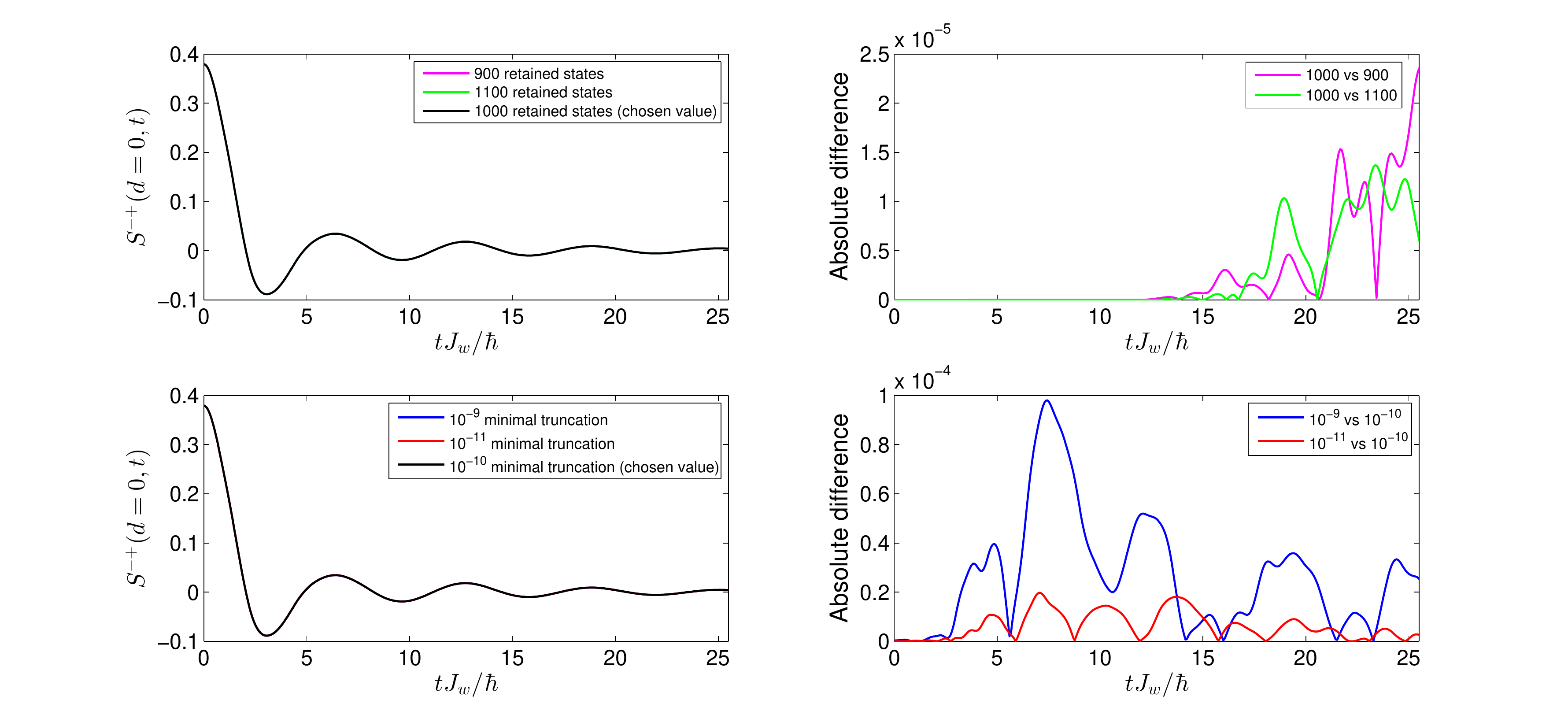}
  \caption{\label{fig:0_4_446k} Case $h=3.016J_w$ (gapless regime) at $T=0.339J_w$. $\left\langle S^-S^+\right\rangle(d=0,t)$ correlations  (left) and the corresponding differences to the chosen value (right) plotted as a function of time, for three different values of retained states (top) and minimal truncation (bottom).}
\end{figure*}

\begin{figure*}
  \centering
  \includegraphics[width=1\textwidth]{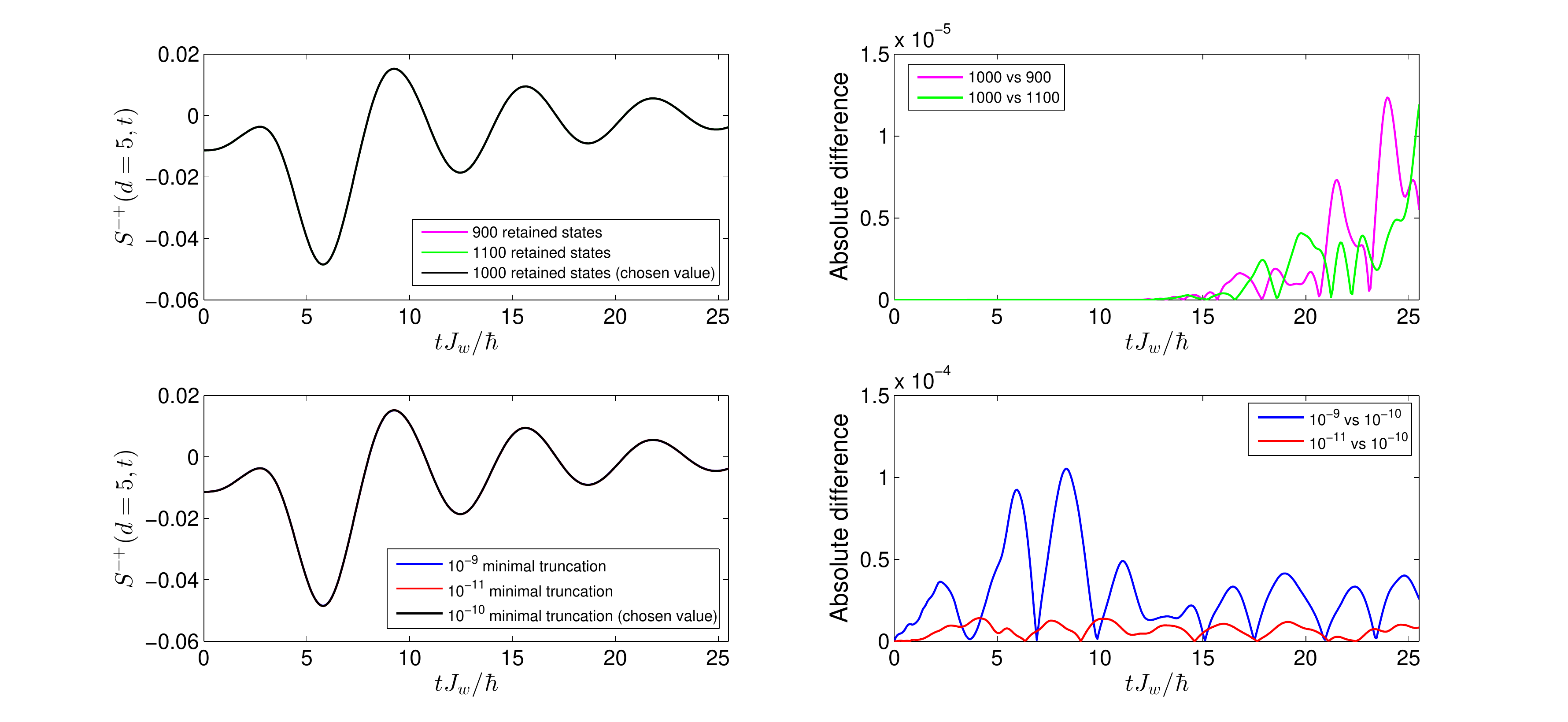}
  \caption{\label{fig:5_4_446k} Case $h=3.016J_w$ (gapless regime) at $T=0.339J_w$. $\left\langle S^-S^+\right\rangle(d=5,t)$ correlations (left) and the corresponding differences to the chosen value (right) plotted as a function of time (left), for three different values of retained states (top) and minimal truncation (bottom).}
\end{figure*}

In this section we show two examples of the convergence checks which we preformed in order to guarantee the validity of our results.
We choose the $\left\langle S^-S^+\right\rangle(d,t)$ correlations and we plot them as a function of time t for $d=0$ (onsite correlations) and $d=5$, and for two different magnetic fields $h=0$ (gapped regime), see Figs.~\ref{fig:0_0k}-\ref{fig:5_0k}, and $h=3.016J_w$ (gapless regime just above $h_{c1}$, see Figs.~\ref{fig:0_4_446k}-\ref{fig:5_4_446k}. The chosen temperature is $T=0.339J_w$. For both values of the magnetic field we consider a larger and a smaller number for the cutoff of the retained states and the minimal truncation error, with respect to the choice of parameters which led us to the results shown in the core of the paper. In all cases we plot as a function of time the correlations themselves (plots on the left) and the absolute difference between the two new choices of parameter (plots on the right). In all cases the correlations plotted in a given figure are substantially indistinguishable and the absolute difference is maximally of the order of $10^{-4}$ such that it is negligible compared to the discretization errors and the error introduced by the finite evolution time considered.


\section{Dynamics of a single excitation in the strong dimerization limit}
\label{app:strongbondpicture}
In the strong dimerization limit, a lot of insight is gained considering a single strong bond weakly coupled to the remaining system. The four-dimensional Hilbert space on a single strong bond is spanned by the four states $\left|s\right\rangle$, $\left|t^+\right\rangle$, $\left|t^0\right\rangle$ and $\left|t^-\right\rangle$. If we put ourselves in a gapped regime, which means $h=0$ or $h>h_{c2}$, the ground state of the system within this picture presents a singlet or a triplet ``+'', respectively, on each strong bond. The application of a spin operator ($S^+$, $S^-$ or $S^z$) can induce  a transition on that specific strong bond to an excited states with a certain energy as detailed below. This excitation can then propagate along the chain since the strong bonds are not totally disconnected from each other ($J_w$ finite). In the following we will detail which transition are possible and what happens for the two cases ($h=0$ or $h>h_{c2}$).

\subsection{Application of the spin operators on a single bond}
The application of the first spin operator of the correlator can excite the system to a higher energy state. These transitions can be used to obtain a crude understanding of the different excitation bands in the correlations. In the following we are giving a table indicating to which states an initial spin state on a bond is transformed by the operation. Taking into account that we always act on the first site of a strong bond we get:
\begin{align}
\label{eq:singlet}
\begin{cases}
S_1^+\left|s\right\rangle\longrightarrow\left|t^+\right\rangle, \\
S_1^z\left|s\right\rangle\longrightarrow\left|t^0\right\rangle, \\
S_1^-\left|s\right\rangle\longrightarrow\left|t^-\right\rangle. 
\end{cases}
\end{align}

\begin{align}
\label{eq:t+}
\begin{cases}
S_1^+\left|t^+\right\rangle\longrightarrow 0, \\
S_1^z\left|t^+\right\rangle\longrightarrow\left|t^+\right\rangle,\\
S_1^-\left|t^+\right\rangle\longrightarrow \left|s\right\rangle+\left|t^0\right\rangle, \\
\end{cases}
\end{align}

\begin{align}
\label{eq:t0}
\begin{cases}
S_1^+\left|t^0\right\rangle\longrightarrow \left|t^+\right\rangle , \\
S_1^z\left|t^0\right\rangle\longrightarrow\left|s\right\rangle, \\
S_1^-\left|t^0\right\rangle\longrightarrow \left|t^-\right\rangle, \\
\end{cases}
\end{align}

\begin{align}
\label{eq:t-}
\begin{cases}
S_1^+\left|t^-\right\rangle\longrightarrow \left|s\right\rangle+\left|t^0\right\rangle, \\
S_1^z\left|t^-\right\rangle\longrightarrow\left|t^-\right\rangle, \\
S_1^-\left|t^-\right\rangle\longrightarrow 0 .
\end{cases}
\end{align}

In the correlations we need to consider in which phase we start in order to know which of the transitions take place and at which energy. This is outlined in the main text and in the subsections below.

\subsection{$h=0$}
In absence of a magnetic field the ground state of the system in the strong dimerization limit is made of singlets on each strong bond. This means that at low temperature mainly the transitions Eq.~\ref{eq:singlet} take place. At zero magnetic field all of them occur at the same energy approximately given by $J_s$. Let us try to see how one of these excitations can hop from one strong bond to the next one. We start by $\left|\varphi^{\alpha}\right\rangle=\left|t^{\alpha}\right\rangle\left|s^{ }\right\rangle$, being $\alpha=+,0$ or $-$, and we apply $H_w=J_w\bold{S}_{j,2}\bold{S}_{j+1,1}$, where the two spin operators act respectively on the second spin of the first bond, and on the first spin of the second bond. What one gets is
\begin{align}
\begin{cases}
H_w\left|\varphi^+\right\rangle=\bold{-\frac{J_w}{4}\left|s^{}\right\rangle\left|t^+\right\rangle}+\frac{J_w}{4}\left|t^+\right\rangle\left|t^0\right\rangle-\frac{J_w}{4}\left|t^0\right\rangle\left|t^+\right\rangle, \\
H_w\left|\varphi^0\right\rangle=\bold{-\frac{J_w}{4}\left|s^{}\right\rangle\left|t^0\right\rangle}+\frac{J_w}{4}\left|t^+\right\rangle\left|t^-\right\rangle-\frac{J_w}{4}\left|t^-\right\rangle\left|t^+\right\rangle, \\
H_w\left|\varphi^-\right\rangle=\bold{-\frac{J_w}{4}\left|s^{}\right\rangle\left|t^-\right\rangle}+\frac{J_w}{4}\left|t^0\right\rangle\left|t^-\right\rangle-\frac{J_w}{4}\left|t^-\right\rangle\left|t^0\right\rangle.
\end{cases}
\end{align}
The most important terms are the three highlighted in bold, which express the hopping of the excitations from one strong bond to the next one. The other terms represent higher energy processes and therefore will be suppressed. From these results it can be seen that the spectrum of the system in all the three cases can be approximated by the one of an isolated particle (the triplet) moving with a tight binding Hamiltonian with hopping matrix element $-J_w/4$.

\subsection{$h>h_{c2}$, $S^{+-}$ correlations}
At very high magnetic field the ground state of the system in the strong dimerization limit is made of triplets ``+'' on each strong bond. The application of $S^-$ induces a transition from a triplet ``+'' to the state $\left|\downarrow\uparrow\right\rangle$, which is a linear combination of $\left|t^0\right\rangle$ and $\left|s\right\rangle$ on a strong bond. For these two excitations, following the same line of reasoning adopted above, one gets that
\begin{align}
\begin{cases}
H_w\left|t^+\right\rangle\left|s^{}\right\rangle=\bold{-\frac{J_w}{4}\left|s^{}\right\rangle\left|t^+\right\rangle}+\frac{J_w}{4}\left|t^+\right\rangle\left|t^0\right\rangle-\frac{J_w}{4}\left|t^0\right\rangle\left|t^+\right\rangle, \\
H_w\left|t^+\right\rangle\left|t^0\right\rangle=\bold{\frac{J_w}{4}\left|t^0\right\rangle\left|t^+\right\rangle}+\frac{J_w}{4}\left|s^{}\right\rangle\left|t^+\right\rangle+\frac{J_w}{4}\left|t^+\right\rangle\left|s^{}\right\rangle.
\end{cases}
\end{align}
The most important resulting terms are again those highlighted in bold, expressing the hopping of the excitations, while the other terms can be neglected since their energies are too high. In this case the spectrum can be approximated by the one of two isolated particles (the triplet ``0'' at high energy, the singlet at low energy) moving with a tight binding Hamiltonian with hopping matrix element respectively $J_w/4$ and $-J_w/4$.

\section{$h=0$, structures at low energy at intermediate temperature}\label{app:lowenstru}
At intermediate temperatures the initial state of the system can contain highly energetic excitations like triplets ``+" or ``-", with fixed initial momentum $q_i$. By invoking conservation of energy, and remembering that $E(q)=-\frac{J_w}{2}\cos(q)$ we get that ($\hbar=1$)
\begin{align}
&E_i(q_i)+\omega=E_f(q_f)\nonumber\\
&\omega=E_f(q_i+q)-E_i(q_i)\nonumber\\
&\omega=-\frac{J_w}{2}\left[\cos(q_i+q)-\cos(q_i)\right]\nonumber\\
&\omega=J_w\sin\left(q_i+\frac{q}{2}\right)\sin\left(\frac{q}{2}\right)
\end{align}
This function has nonzero intensity in the entire region for which $0<\omega/J_w<\sin\left(\frac{q}{2}\right)$, as it can be checked in Fig.~\ref{fig:szsz0}.

\end{document}